\documentclass[12pt]{article}
\usepackage{epsf,amsfonts,amssymb,epsfig,amsmath}
\usepackage{color}
\renewcommand{\text}[1]{#1}

\newcommand{\be}{\begin{equation}}
\newcommand{\ee}{\end{equation}}
\newcommand{\ben}{\begin{displaymath}}
\newcommand{\een}{\end{displaymath}}
\newcommand{\bea}{\begin{eqnarray}}
\newcommand{\eea}{\end{eqnarray}}
\newcommand{\bean}{\begin{eqnarray*}}
\newcommand{\eean}{\end{eqnarray*}}
\newcommand{\nn}{\nonumber \\}
\newcommand{\ba}{\begin{array}}
\newcommand{\ea}{\end{array}}
\newcommand{\bi}{\begin{itemize}}
\newcommand{\ei}{\end{itemize}}

\newcommand{\reef}[1]{(\ref{#1})}

\newcommand{\bbZ}{{\mathbb{Z}}}

\begin{document}

\makeatletter
\renewcommand{\theequation}{\thesection.\arabic{equation}}
\@addtoreset{equation}{section}
\makeatother

\baselineskip 18pt

\begin{titlepage}

\vfill

\begin{flushright}
Imperial/TP/2008/JG/03\\
DESY 08-147
\end{flushright}

\vfill

\begin{center}
  \baselineskip=16pt
 {\Large\bf $AdS_3\times_w(S^3\times S^3\times S^1)$ Solutions\\
 of Type IIB String Theory}
  \vskip 2cm
     Aristomenis Donos$^1$, Jerome P. Gauntlett$^2$ and James Sparks$^3$\\
  \vskip .6cm
     \begin{small}
     $^1$\textit{DESY Theory Group, DESY Hamburg\\
       Notkestrasse 85, D 22603 Hamburg, Germany}
       \end{small}\\*[.6cm]
       \begin{small}
     $^2$\textit{Theoretical Physics Group, Blackett Laboratory, \\
       Imperial College, London SW7 2AZ, U.K.}
       \end{small}\\*[.6cm]
     \begin{small}
     $^2$\textit{The Institute for Mathematical Sciences, \\
       Imperial College, London SW7 2PE, U.K.}
       \end{small}\\*[.6cm]
     \begin{small}
    $^3$\textit{Mathematical Institute, University of Oxford,  \\
        24-29 St Giles', Oxford OX1 3LB, U.K.}
       \end{small}
  \end{center}

\vfill

\begin{center}
\textbf{Abstract}
\end{center}

\begin{quote}
We analyse a recently constructed class of local solutions of type IIB supergravity that
consist of a warped product of $AdS_3$ with a seven-dimensional internal space. In one duality frame
the only other non-vanishing fields are the NS three-form and the dilaton.
We analyse in detail how these local solutions can be extended to 
give infinite families of globally well-defined solutions of type IIB string theory, with the internal space having topology
$S^3\times S^3 \times S^1$ and with properly quantised three-form flux. 
\end{quote}

\vfill

\end{titlepage}
\setcounter{equation}{0}

\tableofcontents

\section{Introduction}
Supersymmetric solutions of string or M-theory that contain $AdS_{d+1}$ factors are dual to
supersymmetric conformal field theories in $d$ spacetime dimensions.
Starting with the work of \cite{Gauntlett:2004zh},
general characterisations of the geometries underlying
such solutions, using $G$-structure techniques \cite{Gauntlett:2002sc,Gauntlett:2002fz},
have been achieved for various $d$ and
for various amounts of supersymmetry \cite{LLM}--\cite{Lunin:2008tf}.
With a few exceptions, mostly with sixteen supersymmetries,
many of these geometries are still
poorly understood, and it has proved
difficult to find explicit solutions.

One notable exception is the class of $AdS_3$ solutions of
type IIB string theory with non-vanishing five-form flux, dual
to $d=2$ conformal field theories with $(0,2)$ supersymmetry,
that were classified in \cite{Kim:2005ez}.
It was shown
that the seven-dimensional internal space has a Killing vector which is
dual to the $R$-symmetry of the dual SCFT. The Killing vector
defines a foliation and the solution is completely determined, locally,
by a K\"ahler metric on the six-dimensional leaf space whose Ricci tensor
satisfies an additional differential condition.
Moreover, a rich set of explicit solutions have been constructed in
\cite{Gauntlett:2006af,Gauntlett:2006qw,Gauntlett:2006ns} and the corresponding
central charges of the dual SCFTs have also been calculated.

More recently, it was understood how to generalise this class of type IIB $AdS_3$ solutions
to also include three-form flux \cite{ajn}.
The solutions are again locally determined by a six-dimensional K\"ahler metric
and a choice of a closed, primitive $(1,2)$-form on the K\"ahler space.
Once again additional explicit solutions were constructed
with the six-dimensional K\"ahler space having a two-torus factor
and the three-form flux being parametrised by a real parameter $Q$.
After two T-dualities on the two-torus it was also shown that
these explicit solutions give type IIB
$AdS_3$ solutions with non-vanishing dilaton and
RR three-form flux only. After an additional $S$-duality
the solutions only involve NS fields.

In \cite{ajn} these explicit solutions were examined in more detail for the special case of $Q=0$.
It was shown that the parameters and ranges of the coordinates could be chosen to give
globally defined supergravity solutions consisting of a warped product of $AdS_3$ with a seven-dimensional
internal manifold that is diffeomorphic to $S^2\times S^3\times T^2$.
It was shown that the solutions, with properly quantised three-form flux, are
specified by a pair of positive coprime integers $p,q$.

The purpose of this paper is to
carry out a similar analysis when we switch on the parameter $Q$. We will find that we are led to infinite
classes of solutions, with the seven-dimensional internal space being diffeomorphic to
$S^3\times S^3\times S^1$ and with properly quantised fluxes.

While the final topology of the solutions is simple, it is not easy to see this in the local coordinates
in which the solutions are presented. When $Q=0$ the $S^2\times S^3$ factor is realised in a manner
very similar to the $Y^{p,q}$ Sasaki-Einstein spaces \cite{Gauntlett:2004yd}. When $Q\ne 0$ one of
the circles in the $T^2$ factor is fibred over the $S^2\times S^3$ and we need to carefully check that
the circle fibration is globally well-defined, leading to $S^3\times S^3$. Furthermore we need to check that
the three-form flux is properly quantised. This is not straightforward since it is not clear
``where'' the two $S^3$ factors are in the local coordinates. After some false starts we
developed a workable prescription for ensuring that the three-form is properly quantised,
as we shall explain.

The plan of the paper is as follows. In section 2, we begin by recalling the local
solutions of \cite{ajn} and then discuss how, after suitable choices of
parameters and periods for the coordinates, the seven-dimensional internal manifold has topology
$S^3\times S^3\times S^1$. We discuss some aspects of the topology in detail, leading
to a prescription for carrying out flux quantisation which is dealt with in section 3.
Our method uses a quotient construction, which is explained in section 2, as well as
explicit coordinate patches. In sections 2 and 3, the solutions depend on a
pair of coprime positive integers $p,q$, the electric three-form flux, $n_1$, the magnetic three-form
flux through each of the two $S^3$ factors, $M_1$ and $M_2$, and the parameter $Q$. For these solutions, it turns out
that $M_1$ and $M_2$ are not independent and are given by $M_1=M(p+q)^2$ and
$M_2=Mq^2$, where $M$ is an integer. We calculate the central charge and
show that it is given by the simple formula
\be\label{introeq}
c=6n_1\frac{(M_1-M_2)M_2}{M_1}~.
\ee
In particular it is independent of $Q$, and since the solutions with $Q\ne 0$ and other parameters fixed are all
smoothly connected with each other, we conclude that when $Q\ne 0$, the parameter $Q$ corresponds to
an exactly marginal deformation in the dual $(0,2)$ SCFT. It is interesting to observe that
the value of the central charge is precisely the same as for the $Q=0$ solutions
studied in \cite{ajn}. However, as we shall explain in section 3.3 taking the limit $Q\to 0$ does not smoothly
lead to the $Q=0$ solutions and so it is not clear whether or not the $Q\ne 0$ solutions correspond to
exactly marginal deformations of the $Q=0$ solutions.


In section 4, we generalise our construction by
making more general identifications on the coordinates, obtaining solutions that involve more parameters.
We show that the central charge has exactly the same form as  in \reef{introeq}, but now, however, the integers
$M_1$ and $M_2$ labelling the three-form flux through the two $S^3$'s are no longer constrained. Thus not all
of these more general solutions correspond to exactly marginal deformations of those that we consider in sections 2 and 3.
We conclude in section 5.

We noted above that the $S^2\times S^3$ factor in the $AdS_3$ solutions constructed in \cite{ajn}, with $Q=0$,
is realised in a similar way to the $Y^{p,q}$ Sasaki-Einstein spaces found in \cite{Gauntlett:2004yd}.
In particular, in both cases the metrics on $S^2\times S^3$ are cohomogeneity one.
Given that the $Y^{p,q}$ metrics can be generalised to cohomogeneity two Sasaki-Einstein metrics $L^{a,b,c}$ on $S^2\times S^3$ \cite{Cvetic:2005ft} (see also \cite{Martelli:2005wy}),
it is natural to suspect that there are analogous $AdS_3$ solutions, with five-form flux only,
with internal space having topology $S^2\times S^3\times T^2$ and with the metric on the $S^2\times S^3$ factor having cohomogeneity two.
This is indeed possible, and moreover it is also possible to find generalisations with non-zero three-form flux
and with the internal manifold having topology $S^3\times S^3\times S^1$. We will present such solutions in appendix C, but we will
leave a detailed analysis of the regularity and flux quantisation conditions for future work.


\section{The $AdS_3$ solutions}


\subsection{The local solutions}
We start with the explicit class of $AdS_3$ solutions of section 4.3 of \cite{ajn}. The string frame
metric is given by
\be\label{sol1}
\frac{1}{L^2}ds^2=\frac{\beta}{y^{1/2}}[ds^2(AdS_3)+ds^2(X_7)]
\ee
where
\bea\label{startingpoint}
ds^2(X_7)&=&\frac{\beta^2-1+2y-Q^2y^2}{4\beta^2}Dz^2+\frac{U(y)}{4(\beta^2-1+2y-Q^2y^2)}D\psi^2
+\frac{dy^2}{4\beta^2y^2 U(y)}\nn
&&+\frac{1}{4\beta^2}ds^2(S^2)+ (du^1-\frac{Qy}{2\beta}[(1-g)D\psi-Dz])^2+(du^2)^2\ ~,
\eea
where $\beta,Q$ are positive constants, $L$ is an arbitrary length scale and
\bea
U(y)=1-\frac{1}{\beta^2}(1-y)^2-Q^2y^2~.
\eea
In addition,
$ds^2(S^2)$ is the standard\footnote{Note that we have rescaled the metric on $S^2$ appearing in \cite{ajn} by a factor of 4.}
metric on a two-sphere, $ds^2(S^2)=d\theta^2+\sin^2\theta d\phi^2$,
and we have defined
\be
D\psi=d\psi+P\ee
with
\be\label{sharon}
dP=\mathrm{Vol}(S^2)=\sin\theta d\theta \wedge d\phi\equiv J~.\ee
Note that $P$ is only a locally defined one-form on $S^2$. In fact,
more precisely, $P$ is a connection one-form
on the $U(1)$ principal bundle associated to the tangent
bundle of $S^2$. The two-form $J$ introduced in (\ref{sharon}) may be regarded as a K\"ahler
form on $S^2$. We also have
\be
Dz=dz-g(y)D\psi
\ee
with
\be
g(y)=\frac{y(1-Q^2y)}{\beta^2-1+2y-Q^2y^2}~.
\ee

The only other non-trivial type IIB supergravity fields are the dilaton and the RR three-form.
The dilaton is given by
\be\label{dial}
e^{2\phi}=\frac{\beta^2}{y}
\ee
while the RR three-form field strength is given by
\bea\label{threeformtdual}
\frac{1}{L^2}F^{(3)}&=&-\frac{1}{4\beta^2}dy\wedge D\psi\wedge Dz
-\frac{y}{4\beta^2}J\wedge Dz+\left[\frac{1-yg}{4\beta^2}\right]J\wedge D\psi\nn
&&+\frac{Q}{2\beta}du^1\wedge [dy\wedge Dz -yJ-(1-g)dy\wedge D\psi]+2 \mathrm{Vol}(AdS_3)~.
\eea
This is closed. After a further $S$-duality transformation we obtain $AdS_3$ solutions with only NS fields non-vanishing,
but we will continue to work with the above solution.

In order to simplify some of the formulae it will be helpful to introduce
\be
Z\equiv1-\sqrt{1+Q^2(\beta^2-1)}~.
\ee
We next change coordinates via
\bea\label{convchange}
dz&=& dw+\frac{2Q\beta}{Z-2}\, dv\nn
du^{1}&=& dv+\, \frac{Q(1-\beta^2)}{2\beta(Z-2)}dw
\eea
to bring the metric to the form
\bea\label{metnewcoords}
ds^{2}(X_7)=
\frac{2(1-Z)(1-\beta^2-yZ)}{(2-Z)(1-\beta^2)}Dv^{2}
+\frac{(1-Z)(2y-yZ-1+\beta^2)}{2\beta^2(2-Z)}Dw^{2}\nn
+\frac{(1-\beta^2)U(y)}{4(1-\beta^2-Zy)(\beta^2-1+2y-Zy)}D\psi^2+\frac{dy^2}{4\beta^2y^2 U(y)}+\frac{1}{4\beta^2}ds^2(S^2)+(du^2)^2\nn
\eea
where
\bea\label{angoneform}
Dv&= & dv-A_{v}\, D\psi\nn
Dw&= & dw-A_{w}\, D\psi\eea
and
\bea
A_{v}&= &
\frac{Q(1-\beta^2)y}{4\beta(1-\beta^2-yZ)}\nn
A_{w}&= &\frac{(2-Z)y}{2(2y-yZ-1+\beta^2)}~.
\eea
The three-form in the new coordinates is given by
\bea\label{tfnewcoords}
\frac{1}{{L}^{2}}F^{(3)}&= & 2\,\mathrm{Vol}\left(AdS_{3}\right)+
\frac{(1-\beta^2)U(y)}{4(1-\beta^2-yZ)(\beta^2-1+2y-yZ)}\, J\wedge D\psi\label{eq:3_form}\\
& -&Dw\wedge\left\{ \frac{(Z-1)(1-\beta^2)}{4\beta^2(Zy-1+\beta^2)}
dy\wedge D\psi
+\frac{y(1-Z)}{4\beta^2} J\right\} \nonumber \\
& -&QDv\wedge\left\{\frac{(1-Z)(1-\beta^2)}{2\beta(Z-2)(-Zy-1+\beta^2+2y)}  dy\wedge D\psi
+\frac{y(1-Z)}{2\beta(2-Z)} J\right\} \nonumber \\
& -&\frac{Q(1-Z)}{\beta(2-Z)}
dy\wedge Dv\wedge Dw\nonumber~. \eea
Finally, we note that
the canonical Killing vector related to supersymmetry is given by
\be
\partial_\psi+\partial_z~.
\ee
In the new coordinates this reads
\bea
\partial_\psi+\frac{(Z-2)}{2(Z-1)}\partial_w
+\frac{Q(\beta^2-1)}{4\beta(Z-1)}\partial_v~.\eea

We now would like to find the restrictions on the parameters $\beta$, $Q$ so that these local solutions
extend to global solutions on a globally well-defined manifold $X_7$. Having achieved that goal, we will
analyse the additional constraints imposed by ensuring that the three-form is properly quantised.
Note that when $Q=0$ the corresponding analysis was carried out in \cite{ajn} and in particular it
was shown that there were infinite classes of solutions, labelled by a pair of positive coprime
integers, $p,q$, with $X_7$ having the topology of
$S^3\times S^2\times T^2$.

Our strategy is to build $X_7$ in stages. The $u^2$ coordinate is taken to paramaterise an $S^1$: for now the period
of $u^2$ is arbitrary but it will later be fixed by flux quantisation. We therefore write $X_7=M_6\times S^1$ with
\be
ds^2(M_6)\equiv \frac{2(1-Z)(1-\beta^2-yZ)}{(2-Z)(1-\beta^2)}Dv^{2}+ds^2(M_5)\ ,
\ee
where
\be
ds^2(M_5)\equiv \frac{(1-Z)(2y-yZ-1+\beta^2)}{2\beta^2(2-Z)}Dw^{2}+ds^{2}(B_4)
\ee
and
\be
ds^2(B_4)\equiv \frac{(1-\beta^2)U(y)}{4(1-\beta^2-Zy)(\beta^2-1+2y-Zy)}D\psi^2+\frac{dy^2}{4\beta^2y^2 U(y)}+\frac{1}{4\beta^2}ds^2(S^2)~.
\ee

We will first analyse $ds^2(B_4)$, showing that, by taking $\psi$ to be a periodic coordinate with period $2\pi$,
$B_4$ is a smooth manifold diffeomorphic to $S^2\times S^2$. We then show
that, by taking  $w$ to be a periodic coordinate with a suitable period, with the parameter $\beta$ fixed
by two relatively prime positive integers $p,q$,  $M_5$ is the total space of a circle fibration over
$B_4$, and has topology $S^3\times S^2$. Here $p$ and $q$ have a topological interpretation
as Chern numbers of the circle bundle over $B_4$.
These steps are familiar from the construction of the Sasaki-Einstein manifolds
$Y^{p,q}$ \cite{Gauntlett:2004yd} (and also for the $Q=0$ solutions studied in \cite{ajn}).
The final step is to show that, by taking $v$ to be periodic with a suitable period,
$M_6$ is the total space of a circle fibration over $M_5$, and has topology $S^3\times S^3$.

It will be useful in the following to observe that the function $U(y)$ is a quadratic function of $y$ with
roots $y_1$ and $y_2$ given by
\bea\label{shelly}
y_1&=&\frac{1-\beta^2}{1+\beta(1-Z)}\nn
y_2&=&\frac{1-\beta^2}{1-\beta(1-Z)}~.
\eea
It will also be useful to know the values
of the functions $A_{w}$ and $A_{v}$ appearing in $ds^2(M_6)$
at $y_1$ and $y_2$. We find
\bea\label{hull}
A_{w}(y_1)&=&\frac{2-Z}{2(1-Z)(1-\beta)}\nn
A_{w}(y_2)&=&\frac{2-Z}{2(1-Z)(1+\beta)}\nn
A_{v}(y_1)&=&\frac{Q(1-\beta)}{4\beta(1-Z)}\nn
A_{v}(y_2)&=&\frac{Q(1+\beta)}{4\beta(1-Z)}~.
\eea


\subsection{$B_4=S^2\times S^2$}\label{4sec}
$B_4$ is parametrised by $\theta$, $\phi$, $y$ and $\psi$.
We take the coordinate $y$ to lie in the interval $y\in [y_1,y_2]$
where $y_i$ are the two
distinct positive\footnote{We need $y$ to be positive to ensure that the warp factor is real.} roots of $U(y)$, given by (\ref{shelly}).
This requires that we demand
\be
0<\beta<1,\qquad 0\le Z<1~.
\ee
We next observe that if we choose the period of $\psi$ to be $2\pi$, then $y,\psi$ parametrise a smooth two-sphere,
with $y$ a polar coordinate and $\psi$ an azimuthal coordinate on the metrically squashed $S^2$ fibre.
In particular, fixing a point on the round two-sphere, one can check that $ds^2(B_4)$ is free from
conical singularities at the poles $y=y_1$ and $y=y_2$. $B_4$ is then a smooth $S^2$ bundle over the round $S^2$.
The transition functions are in $U(1)$, acting in the obvious way on the fibre.
The first Chern number of the $U(1)$ fibration is $-2$ and thus,
as explained in \cite{Gauntlett:2004yd},
$B_4$ is diffeomorphic to $S^2\times S^2$.

We have $H_2(B_4,\bbZ)\cong \bbZ\oplus\bbZ$.
Three obvious two-spheres in $B_4$ are the sections
$\Sigma_1=\{y=y_1\}$ and $\Sigma_2=\{y=y_2\}$, each a
copy of the two-sphere base, and a copy of the
fibre $\Sigma_f$ at some point on the two-sphere base
(for concreteness, say, the north pole $\{\theta=0\}$).
Call the corresponding homology classes $[\Sigma_1]$, $[\Sigma_2]$ and $[\Sigma_f]$, respectively.
We can take $[\Sigma_2]$ and $[\Sigma_f]$ to generate $H_2(B_4,\bbZ)$, but we
note that this is \emph{not} the natural basis of $S^2 \times S^2$. In particular,
the intersections of the 2-cycles are
\bea
[\Sigma_f]\cap [\Sigma_f] = 0, \quad [\Sigma_f]\cap[\Sigma_2] = 1, \quad [\Sigma_2]\cap[\Sigma_2] = 2~.
\eea
The only non-obvious equality above is the last. This follows since the self-intersection
of a 2-cycle in a 4-manifold is equal to the Chern number of the normal bundle.
Similar calculations show that
\be\label{homrel2}
[\Sigma_1]=[\Sigma_2]-2[\Sigma_f].
\ee
Later it will be useful to use a more natural basis given by
$[C_1]=[\Sigma_2]-[\Sigma_f]$ and $[C_2]=[\Sigma_f]$: indeed
one can then check that $[C_1]\cap [C_1]=[C_2]\cap [C_2]=0$ and $[C_1]
\cap[C_2]=1$.

By Poincar\'e duality we have $H_2(B_4,\bbZ)\cong H^2(B_4,\bbZ)$.
Recall that, by definition, the Poincare dual $\eta_{\Sigma}$ of a submanifold $\Sigma\subset M$
satisfies
\bea
\int_\Sigma \omega = \int_M\omega\wedge \eta_{\Sigma}
\eea
for any closed form $\omega$. We introduce the closed two-forms on $B_4$
\bea
\sigma_2&=&\frac{1}{4\pi}J\nn
\sigma_f&=&\frac{1}{2\pi [A_{w}(y_1)-A_{w}(y_2)]}
[(A_{w}(y_2)-A_{w})J-\partial_y(A_{w}) dy\wedge D\psi]~.
\eea
These forms satisfy
\bea\label{dual}
\int_{\Sigma_2}\sigma_2 =\int_{\Sigma_f}\sigma_f = 1, \qquad \int_{\Sigma_2} \sigma_f = \int_{\Sigma_f} \sigma_2 = 0~,
\eea
and one finds that
Poincar\'e duality maps $\Sigma_f\mapsto \sigma_2$ and $\Sigma_2\mapsto \sigma_f+2\sigma_2$.


\subsection{$M_5=S^3\times S^2$}
We next construct $M_5$ as the total space of a circle bundle over $B_4$, by letting $w$
be periodic with period $2\pi l_{w}$, for a suitably chosen
$l_{w}$. We begin by observing from \reef{metnewcoords} that the norm of the
Killing vector $\partial_{w}$ is nowhere-vanishing, and so the size of the
$S^1$ fibre doesn't degenerate anywhere. Recalling that
$Dw=  dw-A_{w}\, D\psi$, we require that $l_{w}^{-1}A_{w}D\psi$
is a connection on a {\it bona fide} $U(1)$ fibration with first Chern
class represented by $(2\pi l_{w})^{-1}d(A_{w}\, D\psi)$.

It is straightforward to first check that
$(2\pi l_{w})^{-1}d(A_{w}\, D\psi)$  is indeed a globally defined two-form on $B_4$.
We next impose that it has integer valued periods:
\bea\label{fperiods}
\frac{1}{2\pi l_{w}}\int_{\Sigma_{2}}d\left(A_{w}D\psi\right) & =&\frac{2}{l_{w}}A_{w}\left(y_{2}\right)=p\nn
\frac{1}{2\pi l_{w}}\int_{\Sigma_{f}}d\left(A_{w}D\psi\right) & =&\frac{1}{l_{w}}\left[A_{w}\left(y_{2}\right)-A_{w}\left(y_{1}\right)\right]=-q,
\eea
where $p, q$ are positive integers. One can then calculate
\be
\frac{1}{2\pi l_{w}}\int_{\Sigma_{1}}d\left(A_{w}D\psi\right)  =\frac{2}{l_{w}}A_{w}
\left(y_{1}\right)=p+2q
\ee
as expected from \reef{homrel2}. We then deduce that
\bea\label{mancity}
\beta=\frac{q}{p+q}
\eea
which, remarkably, is independent of $Q$, and
\be
l_{w}=\frac{2-Z}{p(1-Z)(1+\beta)}~.
\ee
%
With these choices we have that $M_5$ is the total space of a circle bundle with first Chern class given by
\bea
c_1 = p[\sigma_2] - q[\sigma_f]\in H^2(B_4,\bbZ)~.
\eea
As in \cite{Gauntlett:2004yd}, taking $p$ and $q$ to be relatively prime, as we shall henceforth do,
one can show that $M_5$ is simply-connected with $H_2(M_5,\bbZ)\cong \bbZ$. Using Smale's
theorem for five-manifolds \cite{smale}, it follows that $M_5$ is diffeomorphic to $S^3 \times S^2$.

Having constructed $M_5$, it will be useful later to know various topological properties
of this manifold in terms of the coordinate system above.
In the remainder of this subsection we write down explicit generators
for $H^2(M_5,\bbZ)\cong \bbZ$, which will be useful for constructing circle bundles over $M_5$,
and for $H^3(M_5,\bbZ)\cong \bbZ$, which will be useful both for
integration using Poincar\'e duality and also for quantising the three-form flux.
We also find representatives of the generating 2-cycle and 3-cycle in
$H_2(M_5,\bbZ)\cong\bbZ$ and $H_3(M_5,\bbZ)\cong \bbZ$, respectively.

The generator of $H^2(M_5,\bbZ)\cong \bbZ$ may be taken to be the pull-back of the class
\bea\label{lambdaeqn}
\tau = b[\sigma_2] + a[\sigma_f]\in H^2(B_4,\bbZ)
\eea
under the projection
\bea
\pi:M_5\rightarrow B_4~,
\eea
where $a$ and $b$ are (any) integers satisfying
\bea\label{abeqn}
pa+qb=1~.
\eea
These exist and are unique up to $b\rightarrow b+mp$, $a\rightarrow a-mq$, for
any integer $m$, by Bezout's lemma. The non-uniqueness simply corresponds to the fact that the
Chern class $c_1=p\sigma_2-q\sigma_f$ of the circle bundle over $B_4$ is trivial when pulled back
to $M_5$, as is the Chern class of any tensor power of this circle bundle
(the power corresponds to the integer $m$ above).

To see that $\pi^*\tau$ is the generator of $H^2(M_5,\bbZ)$ as claimed, note that, {\it a priori},
$\pi^*\tau$ is necessarily $\delta$ times the generator,
for some integer $\delta\in\bbZ$.
Thus we write $\pi^*\tau=\delta\in H^2(M_5,\bbZ)\cong \bbZ$.
Next note that the circle bundle $\pi$
trivialises over any\footnote{Although $S$ certainly exists, in practice
it is not easy to define such a smooth submanifold in the above coordinate system.}
smooth submanifold $S\subset B_4$ that represents
the cycle
\bea
[S]=q[\Sigma_2]+p[\Sigma_f]~.\eea
This is simply because the first Chern class $c_1$ evaluated on $[S]$ is zero, as one
sees using (\ref{dual}).
Hence we may take a section $s$ of $\pi$ over $S$:
\bea
s:S\rightarrow M_5~.
\eea
This defines a 2-cycle $[s(S)]$ in $H_2(M_5,\bbZ)\cong \bbZ$, which we may take to be $\alpha$
times the generator, for some integer $\alpha$. But then by construction
\bea
\int_{s(S)}\pi^*\tau = \int_S \tau = 1~,
\eea
implying that $\alpha\delta=1$, and thus $\alpha$ and $\delta$ are both $\pm 1$.
Hence $\pi^*\tau$ generates $H^2(M_5,\bbZ)$, and $s(S)$ generates $H_2(M_5,\bbZ)$.

The only other non-trivial homology group is $H_3(M_5,\bbZ)\cong \bbZ$.
There are three natural three-submanifolds of $M_5$, which we call $E_1$, $E_2$ and $E_f$.
These are the restriction of the circle bundle $\pi$ to the submanifolds $\Sigma_1$,
$\Sigma_2$ and $\Sigma_f$ of $B_4$, respectively.
These three-manifolds are all Lens spaces\footnote{See appendix A for some discussion.}.
Indeed, $\Sigma_1$, $\Sigma_2$, $\Sigma_f$
are all two-spheres. The Chern numbers are easily read off from $c_1$ above to be
$p+2q$, $p$ and $-q$. Thus
\bea\label{tophere}
E_1 \cong S^3/\bbZ_{p+2q}, \quad E_2\cong S^3/\bbZ_p, \quad E_f\cong S^3/\bbZ_q~.
\eea
We may take the generator of $H_3(M_5,\bbZ)$ to be
\bea
E = k[E_1]+l[E_f]
\eea
where $k$ and $l$ are (any) integers satisfying
\bea
pk+ql=1~.
\eea
Notice this is the same as (\ref{abeqn}), so one could choose $k=a$ and $l=b$.
A simple way to check this is to note that
the generator has intersection number 1 with $[s(S)]$. One computes
\bea
[s(S)]\cap E = pk+ql=1
\eea
which uniquely identifies $E$ as the generator. We then have
\bea\label{topthere}
[E_1]= p E, \qquad [E_2]=(p+2q)E, \qquad [E_f]=qE~,
\eea
which again can be shown by taking intersection numbers with $[s(S)]$.

Finally, we may also write down a representative $\Phi$ of the generator of
$H^3(M_5,\mathbb{Z})$. By definition this is a closed three-form on $M_5$ that integrates to 1 over $E$.
We choose
\begin{align}\label{m5threegen}
\Phi=\frac{1}{\left(2\pi\right)^{2}l_{w}^{2}} &
\left\{ Dw\wedge \left[\left(A_{w}\left(y_{1}\right)+A_{w}\left(y_{2}\right)-A_{w}\left(y\right)\right)J-
\partial_yA_w dy\wedge D\psi\right]\right.\nn
&\left.-\left[A_{w}^2(y)-A_w\left(y\right)\left(A_{w}\left(y_{1}\right)+A_{w}\left(y_{2}\right)\right)+A_{w}\left(y_{1}\right)A_{w}\left(y_{2}\right)\right]\, J\wedge D\psi\right\} .
\end{align}
The three-form $\Phi$ is Poincar\'e dual to the non-trivial two-cycle in $M_5$.


\subsection{$M_6=S^3\times S^3$}
We now construct $M_6$ as a circle bundle over $M_5$.
Since $H^2(M_5,\bbZ)\cong \bbZ$, such circle bundles are determined,
up to isomorphism, by an integer. Since $M_5\cong S^3 \times S^2$,
taking this integer to be $1$ (or $-1$) gives a total space
$M_6\cong S^3 \times S^3$. Taking the Chern number to be $n$ would instead lead to an $M_6$ with
$\pi_1(M_6)\cong \bbZ_{n}$, which we may always lift to the simply-connected cover with
$n=\pm 1$. So, we will do this. However, as we shall see later, in fixing the three-form
flux quantisation it will be helpful to consider such quotients of $M_6$.

Observe from \reef{metnewcoords} that the norm of the
Killing vector $\partial_{v}$ is nowhere-vanishing, and so the size of the
$S^1$ fibre doesn't degenerate anywhere. The period
of $v$ is taken to be $2\pi l_{v}$, where $l_{v}$ will be
fixed shortly. Recalling that
$Dv=  dv-A_{v}\, D\psi$, we require that $l_{v}^{-1}A_{v}D\psi$
is a connection on a $U(1)$ fibration with first Chern
class represented by $(2\pi l_{v})^{-1}d(A_{v}\, D\psi)$.
It is straightforward to check that
$(2\pi l_{v})^{-1}d(A_{v}\, D\psi)$  is a globally defined two-form on $M_5$.
We next impose that it has unit period. To do this we would like to integrate
$(2\pi l_{v})^{-1}d(A_{v}\, D\psi)$ over
a smooth submanifold in the same homology class as $s(S)$, the generator of
$H_2(M_5,\bbZ)$. However, as we have already noted, finding such a smooth
submanifold is not so easy. Luckily, we can use Poincar\'e duality to
calculate the period instead. Recalling that $[\Phi]$ is Poincar\'e dual to $[s(S)]$, we
 demand that
\bea\label{pdtrick}
\frac{1}{2\pi l_{v}}\int_{s(S)} d(A_{v}D\psi)&=&
\frac{1}{2\pi l_{v}}\int_{M_5} d(A_{v}D\psi)\wedge \Phi\nn
&=&\frac{2}{l_v l_w}[A_v(y_2)A_w(y_1)-A_v(y_1)A_w(y_2)]\nn
&=&\frac{1}{l_v}[2qA_v(y_2)-p(A_v(y_1)-A_v(y_2))]=1,
\eea
so that the circle bundle has Chern number 1, which
can be achieved by setting
\be\label{manu}
l_{v}=\frac{Q(p+q)}{1-Z}~.
\ee

Let us denote this circle bundle over $M_5$ by $L$, with corresponding projection
\be
\Pi:M_6\to M_5~.\ee
Recalling that the generator of $H^2(M_5,\bbZ)$ may be taken to be the pull-back of
$\tau$ in (\ref{lambdaeqn}) under the projection $\pi:M_5\to B_4$, we see that $L$
may be regarded as the pull-back of the circle bundle
$L_{\tau}$ over $B_4$ with first Chern class given by $\tau\in H^2(B_4,\bbZ)$. We
write this as $L=\pi^*L_{\tau}$.

Since $M_6\cong S^3 \times S^3$, it follows that the only non-trivial homology group is
$H_3(M_6,\bbZ)\cong \bbZ\oplus\bbZ$.
The two generators are clearly the two copies of $S^3$, at a fixed point on the other copy.
However, because of the way we have constructed $M_6$ above, it is not easy to see the
diffeomorphism of $M_6$ with $S^3\times S^3$ explicitly.
Nevertheless, we observe that one three-cycle is represented by the total space of the circle bundle $L$ over the $S^2$ in $M_5\cong S^3\times S^2$.
Since $s(S)$ is homologous to the
$S^2$ in $M_5$, it follows\footnote{Being homologous in $M_5$ means there is a
three-dimensional chain in $M_5$ with boundary $s(S)-S^2$. By taking the total space
of $L$ over this chain, one obtains a chain in $M_6$ with boundary given by the total
space of $L$ over $s(S)-S^2$.} that taking the total space of $L$ over both submanifolds gives
homologous three-submanifolds of $M_6$, which is the total space of $L$.
Thus the total space of the $L$ circle bundle
over $s(S)$ is one of the generators of the homology of $M_6$. It should be pointed out, though, that
finding a smooth representative of this generator is not straightforward.
For the other generator, the obvious thing
to try is to take a representative for $E$, which afterall is represented by $S^3\subset M_5$,
and then try to take a section of $\Pi$ over this representative. However,
unfortunately just because two submanifolds are homologous in $M_5$, with $L$
trivial over one of them, this does not necessarily guarantee
that the circle bundle $L$ is trivial over the other submanifold\footnote{As a simple
example, consider the five-manifold $T^{1,1}\cong S^2\times S^3$, which recall
is naturally a circle bundle over $S^2\times S^2$. For our two three-submanifolds we take
a contractible $S^3$, say the equatorial $S^3$ on a contractible $S^4$ that links a point,
and the ``diagonally embedded'' Lens space $S^3/\bbZ_2$. Since $T^{1,1}$ is
a circle bundle over $S^2\times S^2$, we may describe the latter three-submanifold more precisely
as the restriction of this circle bundle to the diagonal $S^2$ in $S^2\times S^2$, which
is the easiest way to see that the topology is indeed $S^3/\bbZ_2$. Both
three-cycles are trivial -- to see this for the latter construct the generator of
$H^3(T^{1,1},\bbZ)$ and integrate over the three-cycle. However, if we pull back
the complex line bundle $\mathcal{O}(1,0)_{S^2\times S^2}$ with winding numbers 1 and 0
on $S^2\times S^2$ to $T^{1,1}$, this is trivial over the $S^3$ but non-trivial
over the contractible $S^3/\bbZ_2$ (the latter follows using arguments similar to those in appendix B).}.
So, we cannot necessarily do this. An additional observation is that,
while a section of $\Pi$ exists over $E$, it does not exist, in general, over
the submanifolds $E_1$, $E_2$ and $E_f$, as we explain in the appendix.

In order to carry out the flux quantisation of the three-form in the supergravity solutions,
we need a prescription to integrate three-forms over a basis of $H_3(M_6,\bbZ)$. The comments
in the last paragraph indicate that this is not as straightforward as it might seem.
Our approach, employing a quotient construction\footnote{We thank Dominic Joyce for suggesting this
approach.}, will be explained in the next subsection.


\subsection{A quotient of $M_6$ and integral three-forms}\label{quotientsection}

In this section we want to explain how considering the periods
of the three-form on the quotient $\hat M_6=M_6/\bbZ_{(p+q)q}$ leads
to a practical procedure for ensuring that a three-form, such as the suitably
normalised RR three-form, has integral periods.

In order to obtain more insight into the topology of $M_6$, it will be helpful
to think of it as a group manifold,
\bea
M_6 =
S^3\times S^3\cong SU(2)\times SU(2)~,\eea
and observe that taking the quotient by the maximal torus $T^2\subset SU(2)\times SU(2)$ leads to
$B_4$:
\bea
M_6/T^2 = S^2\times S^2 = B_4.
\eea

Now, recall that we constructed $M_5$ as the total space of a circle bundle
over $B_4$ with winding numbers $p$ and $-q$ over $\Sigma_2$ and $\Sigma_f$,
respectively. With respect to the natural basis $[C_1]=[\Sigma_2]-[\Sigma_f]$ and $[C_2]=[\Sigma_f]$ of $B_4\cong S^2\times S^2$
introduced in section \ref{4sec}, we thus have Chern numbers
\bea
\int_{[C_1]} c_1 = p+q, \qquad \int_{[C_2]} c_1 = -q~.
\eea
In this section we make the $U(1)$ fibration structure of $M_5$ explicit in the notation
by denoting the latter as $M_5(p,q)$.

The key observation is that
we may realise $M_5(p,q)$ as a quotient by the $U(1)$ subgroup of $T^2$
with charges $(q,p+q)$, as illustrated in the following diagram:
\bea\label{diagram}
\begin{array}{ccccc}& U(1)_{q,p+q} & & & \\
\ \ \ \swarrow & & \searrow & & \\
T^2 & \hookrightarrow & \ \ \ \ M_6 & \ \ \rightarrow & \quad B_4 \\
\downarrow & & \ \ \ \ \downarrow & &  \\
T^2/U(1)_{q,p+q} & \hookrightarrow & \ \ \ \ M_5(p,q) & \ \ \rightarrow & \quad B_4\end{array}~.
\eea
To see this more explicitly, we introduce Euler angles, $\psi_1,\theta_1,\phi_1$
and $\psi_2,\theta_2,\phi_2$ for each of the two $SU(2)$ factors.
We also introduce the corresponding left-invariant one-forms
$\sigma^{\alpha}_i$ for each factor, respectively, where $\alpha=1,2$; $i=1,2,3$.  Thus
\bea
\sigma^{\alpha}_1 &=& \cos\psi_\alpha d\theta_\alpha + \sin\theta_\alpha\sin\psi_\alpha d\phi_\alpha\nn
\sigma^{\alpha}_2 &=& -\sin\psi_\alpha d\theta_\alpha + \sin\theta_\alpha\cos\psi_\alpha d\phi_\alpha\nn
\sigma^{\alpha}_3 &=& d\psi_{\alpha}+\cos\theta_\alpha d\phi_\alpha~.
\eea
Now $\psi_1,\psi_2\in [0,4\pi)$ parametrise the $T^2$.
The $U(1)_{q,p+q}$ circle action is then given explicitly by
\bea\label{circact}
(\psi_1,\psi_2)\mapsto (\psi_1 + q \psi, \psi_2+(p+q)\psi)
\eea
where $\psi\in [0,4\pi)$ parametrises the circle subgroup.
If we introduce coordinates $\tilde{v},\tilde{w}$
defined by
\bea
\tilde{v}=-\frac{1}{q}\psi_1,\qquad \tilde{w}=(p+q)\psi_1-q\psi_2
\eea
then the $T^2$ is parametrised by taking $\tilde{v},\tilde{w}\in [0,4\pi)$.
In these coordinates the $U(1)_{q,p+q}$ circle action reads
\bea
(\tilde{v},\tilde{w})\mapsto (\tilde{v}-\psi,\tilde{w})
\eea
and hence $\tilde{w}$ parametrises the circle $T^2/U(1)_{q,p+q}$.
The globally defined connection one-form on the total space of the
circle bundle on the bottom line of (\ref{diagram})
is given by
\bea\label{angexp}
\eta &=& \frac{1}{2}((p+q) \sigma^1_3 -q \sigma^2_3)\nn
&=&\frac{1}{2}( d\tilde{w}+(p+q)\cos\theta_1d\phi_1-q\cos\theta_2d\phi_2)~.
\eea
We can define two natural copies of $S^2$ in $B_4$ to be $C_1$ and $C_2$, which are round $S^2$s
at the north pole of the other. So, $C_1=\{\theta_2=0\}$, $C_2=\{\theta_1=0\}$.
We observe that \reef{angexp} gives rise to Chern numbers
$p+q$ and $-q$ for $C_1$ and $C_2$, respectively, as required for $M_5(p,q)$.

Let us denote the total space over each sphere $C_1$ and $C_2$ in $M_5(p,q)$ to be $F_1$ and $F_2$, respectively.
Then by following similar arguments as in \reef{tophere}--\reef{topthere} we deduce that
\bea\label{isoFs}
F_1 \cong S^3/\bbZ_{p+q}, \qquad F_2 \cong S^3/\bbZ_{q}
\eea
and also the homology relations
\bea
[F_1] = (p+q) [S^3], \qquad [F_2] = q [S^3]~.
\eea
In fact one can see (\ref{isoFs}) rather explicitly from the above quotient construction.
We define $W_1\cong S^3$ and $W_2\cong S^3$
to be the two natural copies of $S^3$ in $M_6$ given by $W_1=\{\theta_2=0,\psi_2=0\}$,
$W_2=\{\theta_1=0,\psi_1=0\}$. Consider now $\{\theta_2=0\}\subset M_6$. This is
\bea
W_1\times S^1 \cong S^3\times S^1~,
\eea
where the $S^1$ is parametrised by $\psi_2$.
When we take the quotient by the $U(1)_{q,p+q}$ circle action \reef{circact}
we may set $\psi_2=0$. However, there is then a remaining gauge freedom given by setting
\bea
\psi = \frac{4\pi k}{p+q}~,
\eea
with $k=1,\ldots,p+q$, since this also fixes $\psi_2=0$. This then acts
on $\psi_1$, which is the Hopf fibre
of $W_1$ realised as an $S^1$ bundle over $S^2$, and we see explicitly that $F_1\cong S^3/\bbZ_{p+q}$.
A similar argument applies to $F_2$.

We next observe that
\bea
\Phi=\frac{1}{8\pi^2}[(p+q) \eta\wedge \sigma^1_1\wedge\sigma^1_2 +q\eta\wedge \sigma^2_1\wedge\sigma^2_2]
\eea
is a closed globally defined three-form on $M_5(p,q)$.
We see explicitly that
\bea
\int_{F_1}\Phi = \frac{p+q}{8\pi^2}\int_{F_1}\eta\wedge\sigma^1_1\wedge\sigma^1_2 = p+q~.
\eea
which shows that $\Phi$ generates $H^3(M_5(p,q),\bbZ)$.

Next it is convenient to define $\hat{M_6}$ to be
\bea
\hat{M_6} = M_6/\bbZ_{(p+q)q}
\eea
where we embed $\bbZ_{(p+q)q}$ along $U(1)_{q,p+q}$. This defines a
quotient
\bea
f:M_6\rightarrow \hat{M_6}~.
\eea
The action on the Euler angles is
\bea\label{PQaction}
(\psi_1,\psi_2)&\mapsto &\left(\psi_1+\frac{4\pi k q}{(p+q)q},\psi_2+\frac{4\pi k (p+q)}{(p+q)q}\right)\nonumber\\
&=& \left(\psi_1+\frac{4\pi k}{p+q},\psi_2+\frac{4\pi k}{q}\right)~.
\eea
Here $k=1,\ldots,(p+q)q$. This realises the $\bbZ_{(p+q)q}$ action as a $\bbZ_{p+q}\times \bbZ_{q}$
action (the groups are isomorphic as $p+q$ and $q$ are coprime) and we have
\bea\label{productLens}
\hat{M_6} \cong (S^3/\bbZ_{p+q})\times (S^3/\bbZ_{q})~.
\eea
In terms of $\tilde{v}$,$\tilde{w}$ we have
\be
(\tilde{v},\tilde{w})\mapsto (\tilde{v}-\frac{4\pi k}{(p+q)q},\tilde{w})~.
\ee
Thus on $\hat M_6$ we can introduce a new coordinate $\hat v =(p+q)q \tilde{v}$ with period $4\pi$
and we also have
\bea
\hat{M_6}\cong (S^3/\bbZ_{(p+q)q}) \times S^3~,
\eea

A key point is that the $\hat{v}$ circle bundle trivialises over both
$F_1$ and $F_2$. One way to see this is to observe that the $\hat{v}$ circle bundle
has first Chern class being $q(p+q)$ times the generator of $H^2(M_5(p,q),\bbZ)$ and then
following the arguments in the appendices. We can also see this directly.
Consider again
\bea
W_1 \times S^1
\eea
where the $S^1$ is coordinatised by $\psi_2$. The action of $\bbZ_{p(p+q)}$ is
given by (\ref{PQaction}). We first set $k=nq$, with $n=1,\ldots,p+q$.
This defines a $\bbZ_{p+q}$ subgroup that acts trivially on $\psi_2$, but acts
non-trivially on $W_1$, with quotient $W_1/\bbZ_{p+q}\cong S^3/\bbZ_{p+q}=F_1$.
We may then set $k=1,\ldots, q$ in the identification. This now acts
trivially on $W_1/\bbZ_{p+q}$, but acts non-trivially on $S^1$ to give
$S^1/\bbZ_{q}\cong S^1$. This shows explicitly that
\bea
(W_1 \times S^1)/\bbZ_{(p+q)q}\cong F_1\times S^1
\eea
which in turn shows that the $\hat{v}$ bundle restricted to $F_1$ is trivial,
as it is manifestly a product. Obviously, similar reasoning applies\footnote{A point we shall return to later,
in passing, is that the above arguments show that for the quotient
$M_6/\bbZ_{p+q}$ the corresponding circle bundle trivialises over $F_1$, while
for $M_6/\bbZ_{q}$ it trivialises over $F_2$. We consider
$M_6/\bbZ_{(p+q)q}$ as it trivialises over both.}
to $F_2$.

Let us now define $V_1$ and $V_2$ to be the obvious 2 factors of $\hat M_6$
in (\ref{productLens}).
Because of the discrete identification (\ref{PQaction}),
$W_1$ is a $(p+q)$-fold cover of $V_1$, and $W_2$ is a $q$-fold cover of $V_2$.
Thus for any three-form $\Psi$ on $\hat{M_6}$ we have
\bea\label{cover}
\int_{W_1} f^*\Psi &=& (p+q) \int_{V_1} \Psi\nn
\int_{W_2} f^*\Psi &=& q\int_{V_2}\Psi~.
\eea
Here $f^*\Psi$ is obtained by simply replacing $\hat v$ in $\Psi$ with $(p+q)q \tilde{v}$.

For example, if we let $\Pi:M_6\to M_5(p,q)$ be the projection for the fibration in the second
column in \reef{diagram}, then $\Pi^*\Phi$ is a three-form on $M_6$ that is invariant under $f$
(it has no dependence on the coordinate $\tilde{v}$).
It is therefore obviously the pull-back of a three-form on the quotient $\hat M_6$,
and hence we may use \reef{cover} to calculate
\bea\label{tip}
\int_{W_1}\Pi^*\Phi &=& (p+q)^2 \nn
\int_{W_2}\Pi^*\Phi &=& q^2.
\eea


Finally, we are in a position to provide our prescription for
quantising the RR flux. We first observe that while we may take $C_2=\Sigma_f$,
we cannot quite take $C_1$ to be $\Sigma_2\cup(-\Sigma_f)$, because the two submanifolds intersect at a point
and we don't have a smooth submanifold. We may remedy this by cutting out a small neighbourhood
of the intesection point and gluing in a cylinder. This results in a two-sphere, which
we can take to be $C_1$. We may then identify
\bea
F_1 &=& E_2\cup (-E_f) \nn
F_2 &=& E_f \cong S^3/\bbZ_q~,
\eea
with the understanding that $F_1$ is to be smoothed out into $S^3/\bbZ_{p+q}$, rather
than the union of $S^3/\bbZ_{q}$ with $S^3/\bbZ_{p}$ over the circle where they intersect.
As we have shown, on $\hat M_6$ the $\hat v$ circle fibration trivialises over $F_1$ and $F_2$,
and hence we may take sections giving submanifolds $V_1$ and $V_2$.
The correct quantisation condition for an integral three-form on $M_6$ (such as our appropriately
normalised RR three-form), in a workable form, is then
given by (\ref{cover}), where the integrals over $W_1$ and $W_2$ are integers $M_1$, $M_2$.


\section{Flux Quantisation and the Central Charge}
In order to obtain a good solution to string theory, we need to
impose that both the electric and magnetic
RR three-form charges are properly quantised.
In this section we analyse this in detail
and then derive the central charge of the corresponding $(0,2)$ SCFT.
We conclude the section with a discussion of taking the limit
$Q\to 0$ in our solutions and the relationship to 
the $Q=0$ solutions of \cite{ajn}.

\subsection{Electric and magnetic charges}
For the electric charge
we require
\bea\label{fqnc}
n_1=\frac{1}{(2\pi l_s)^6 g_s}\int_{X_7} *F^{(3)} \in \bbZ~.
\eea
Since
\be
\frac{1}{L^6}*F^{(3)}=\frac{(Z-1)}{8(Z-2)\beta^2 y^2}J\wedge dy\wedge D\psi
\wedge Dw\wedge d v \wedge du^2
+\mathrm{Vol}(AdS_3)\wedge(\dots)
\ee
we have
\bea
n_1=\left(\frac{L}{l_s}\right)^6\frac{1}{g_s8 \pi^2}\frac{Q(p+q)^5}{p^2q(p+2q)^2}
\Delta u^2~,
\eea
which we interpret as fixing the period, $\Delta u^2$, of the $u^2$ circle.

We next turn to the magnetic three-form charge. We require that
\be
\frac{1}{(2\pi l_s)^2g_s}\int_W F^{(3)}\in \bbZ
\ee
when integrated over any three-cycle $W\subset X_7=M_6\times S^1$.
The relevant three-cycles are in $M_6$, and so the quantisation
condition amounts to quantising the restriction of
$F^{(3)}$ to $M_6$ at a point on the $S^1$ coordinatised by $u^2$.
In the previous subsection we gave a prescription for
performing such integrals by instead calculating integrals on submanifolds
of the quotient space $\hat M_6$.
In the next subsection we
will calculate these integrals by introducing explicit
coordinate patches. This will illuminate and confirm many of our observations
about the topology in the previous section. Furthermore, the techniques will be essential
for the generalisation that we consider in section 4.

In the present case, however, there is a much simpler way to impose flux quantisation.
The key observation is that, remarkably, the relevant part of $F^{(3)}$ is
in the same cohomology class as\footnote{Here we are not distiguishing between $\Phi$ and $\Pi^*\Phi$.}
$\Phi$. Indeed
we have
\bea\frac{1}{L^2}F^{(3)}-2\mathrm{Vol}(AdS_3)
=\frac{(2\pi)^2l_w(1-Z)}{(Z-2)q\beta}\Phi
+d\left\{K_1Dv\wedge Dw +K_2Dw\wedge D\psi\right\}
\eea
where
\bea
K_1&=&\frac{Q(-2y+3Zy-Z^2y+1-\beta^2-Z+Z\beta^2)}{\beta(Z-2)^2}\nn
K_2&=&\frac{(1-\beta^2)(1-Z)U(y)}{(-1+\beta^2+2y-Zy)(-1+\beta^2+Zy)(2-Z)}~.
\eea
Note in particular that the function $K_2$ vanishes at $y_1$ and $y_2$, ensuring that
the two-form $K_1Dv\wedge Dw +K_2Dw\wedge D\psi$ is globally defined.
We thus conclude that
\be
\frac{1}{(2\pi l_s)^2g_s}\int_W F^{(3)}=-\frac{{L}^2}{l_s^2 g_s}\frac{(p+q)^2}{pq^2(p+2q)}
\int_W \Phi~.
\ee
Furthermore, we have already calculated the periods of $\Phi$ (more precisely, $\Pi^*\Phi$) over
a basis of three-cycles on $M_6$ in \reef{tip}. We find that if the length scale
is taken to be
\be\label{resone}
\frac{{L}^2}{l_s^2 g_s}=\frac{pq^2(p+2q)M}{(p+q)^2}
\ee
for some positive integer $M$, then
\bea\label{restwo}
M_1&\equiv& \frac{-1}{(2\pi l_s)^2 g_s}\int_{W_1} F^{(3)} = M(p+q)^2\nn
M_2&\equiv& \frac{-1}{(2\pi l_s)^2 g_s}\int_{W_2} F^{(3)} = Mq^2~.
\eea

We may now calculate the central charge of the dual SCFT.
It is given by \cite{Brown:1986nw}
\be\label{bh}
c=\frac{3R_{AdS_3}}{2G_{(3)}}
\ee
where $G_{(3)}$ is the three-dimensional Newton's constant and
$R_{AdS_3}$ is radius of the $AdS_3$ space.
In our conventions the type IIB supergravity Lagrangian
has the form
\be\label{bhconv}
\frac{1}{(2\pi)^7g_s^2l_s^8}{\sqrt{-\det g}}e^{-2\phi}R+\dots
\ee
and after a short calculation we find
\bea
c&=&6n_1\left(\frac{L}{l_s}\right)^2\frac{1}{g_s}\nn
&=&6n_1\frac{pq^2(p+2q)M}{(p+q)^2}=6n_1\frac{(M_1-M_2)M_2}{M_1}\ .
\eea
This result is independent of $Q$ and since the solutions with
$Q\ne 0$ and all other parameters fixed are smoothly connected, we conclude that the parameter $Q$ corresponds
to an exactly marginal deformation in the dual SCFT. It is interesting to observe that
the central charge is the same as that of the $Q=0$ solutions studied in \cite{ajn}.
However, as we shall explain in section 3.3, taking the limit $Q\to 0$ does not
smoothly give the $Q=0$ solutions of \cite{ajn} and so it is not clear whether or
not the $Q\ne 0$ solutions are exactly marginal deformations of those with $Q=0$.


\subsection{Computing periods using coordinate patches}

In this subsection we directly compute the flux of $F^{(3)}$ through
the two three-cycles of $M_6$ using coordinate patches. This provides
a nice cross-check on various calculations carried out so far. Furthermore,
we will use this method in the next section when we construct more general
type IIB string theory solutions -- there we will not be able to
use the approach
in the last subsection since the three-form flux will no longer be in the same
cohomology class as $\Pi^*\Phi$.

Recall from section \ref{quotientsection} that
instead of considering the circle bundle $L$ over $M_5$
with total space $M_6$ we should
consider the circle bundle $\hat{L}=L^{(p+q)q}$ with total space
$\hat M_6=M_6/\bbZ_{(p+q)q}$. This is useful since $\hat{L}$ trivialises over
both the submanifolds $F_1$, a smoothed out version of $E_2\cup -E_f$,
and $F_2\equiv E_f$ of $M_5$. We may thus take sections of $\hat{L}$ over
these submanifolds to obtain submanifolds
$V_1$ and $V_2$ of $\hat{M}_6$. Then the quantisation of
the three-form flux on $M_6$, through the two three-cycles
$W_1$, $W_2$, is related to that on $\hat{M}_6$ via the general formulae (\ref{cover}).

In particular, this procedure involves trivialising
the circle bundle $\hat{L}$ over $F_1$ and $F_2$. Concretely,
this means that the corresponding connection one-form is
a globally-defined one-form over $F_1$ and $F_2$. However,
to see this requires carefully covering the manifold
with coordinate patches, so that the connection form
is represented by a globally defined one-form on each patch, and then gluing
these forms together on overlaps using $U(1)$ transition
functions. Only when one has picked a gauge where
the connection one-form is globally defined on $F_1$, $F_2$ can one
then represent a section by taking the (appropriately gauge transformed)
$v$ coordinate to be constant in the
three-form flux $F^{(3)}$. This might sound
overly-technical, but if one does not follow this
carefully one obtains incorrect periods for the flux.

We begin by covering $M_5$ with 4 coordinate patches: $U_{1N}$, $U_{2N}$,
$U_{1S}$, $U_{2S}$. Here, for example, $U_{1N}$ is defined
by removing $\{y=y_2\}$ and $\{\theta=\pi\}$, while
$U_{1S}$ is defined by removing $\{y=y_2\}$ and
$\{\theta=0\}$. On $B_4$ the points we remove in each
case are
two $S^2s$ that intersect over a point. It follows
that, regarded as defining subsets of $B_4$,
the above conditions give 4 patches diffeomorphic
to $\mathbb{R}^4$. On $M_5$ we
thus obtain patches diffeomorphic to $S^1\times \mathbb{R}^4$, with the $S^1$
in each patch parametrised by a coordinate $w_{1N}, w_{2N}, w_{1S}, w_{2S}$,
respectively.

Recall that $B_4$ is constructed as an $S^2$ bundle over
$S^2$, where the fibre $S^2$ has poles $\{y=y_1\}$,
$\{y=y_2\}$. Removing these, one can define a global one-form:
\bea
D\psi&=&D\psi_N = d\psi_N + (1-\cos\theta)d\phi\nn
&=&D\psi_S = d\psi_S - (1+\cos\theta)d\phi~.
\eea
The corresponding space is an $I\times S^1$ bundle
over $S^2$, where $I=(y_1,y_2)$ is an open interval, and
the circle $S^1$ is parametrised by $\psi_N$ and $\psi_S$,
each with period $2\pi$.
Here the first expression is valid on the complement
of the south pole $\{\theta=\pi\}$, while the second is
valid on the complement of the north pole $\{\theta=0\}$.
This is because the azimuthal coordinate $\phi$ degenerates
at the poles of the base $S^2$. On the overlap one has
\bea
\psi_S -\psi_N = 2\phi
\eea
which shows that the $S^1$ bundle has Chern number $-2$.
This is because the connection form is locally $\cos\theta d\phi$, and so has curvature form
$-\sin\theta d\theta\wedge d\phi$, which integrates to $-2\cdot 2\pi$
over the $S^2$.
It is important that $D\psi$ is \emph{not} defined
at $\{y=y_i\}$, since these are coordinate singularities.

Recalling \reef{angoneform}, we next define the global one-form on $M_5$:
\bea\label{dubs}
Dw &=&Dw_{1N} = dw_{1N} + A_{w}(y_1)d\psi_N - A_{w}D\psi_N\nn
&=&Dw_{2N} = dw_{2N} + A_{w}(y_2)d\psi_N - A_{w}D\psi_N\nn
&=&Dw_{1S} = dw_{1S} + A_{w}(y_1) d\psi_S - A_{w}D\psi_S\nn
&=&Dw_{2S} = dw_{2S} + A_{w}(y_2) d\psi_S - A_{w}D\psi_S~.
\eea
These are defined on the 4 patches $U_{1N}$, $U_{2N}$,
$U_{1S}$, $U_{2S}$, respectively. Take, for example,
$Dw_{1N}$. $\psi_N$ is a coordinate on the complement
of the south pole of the base $S^2$, although it degenerates
at $y=y_1$. However, at $y=y_1$ we have
\bea
Dw_{1N}\mid_{\{y=y_1\}} = dw_{1N} - A_{w}(y_1) (1-\cos\theta)d\phi~.
\eea
and we see that $w_{1N}$ is indeed a good coordinate on the $S^1$ of $U_{1N}\cong
S^1\times \mathbb{R}^4$. The period of all the $w$ coordinates
above is $2\pi l_{w}$.

One can immediately see the fibration structure of the
$w$ circle bundle, with total space $M_5$, from the above formulae.
For example, on the overlap region where both are defined, we have
\bea\label{qtwistnew}
\frac{1}{l_{w}}(w_{2N}-w_{1N}) = q\psi_N~.
\eea
In particular, restricting to $\{\theta=0\}$, which is
$E_f$, we see that the circle bundle has Chern number $-q$
and thus $E_f\cong S^3/\bbZ_q$. Similarly,
\bea\label{ptwistnew}
\frac{1}{l_{w}}(w_{2S}-w_{2N}) = -p\phi
\eea
showing that the Chern number over $E_2=\{y=y_2\}$ is
$p$, thus proving that $E_2\cong S^3/\bbZ_p$.

In each of the patches we define the connection one-form that appears in the $v$ circle fibration
over $M_5$ to give $M_6$. Recalling \reef{angoneform} we write $Dv\equiv dv-A'$ and define
\bea\label{lambdas}
A'_{1N} &=& -A_{v}(y_1)d\psi_N+A_{v}  D\psi_N+l_v\lambda_{1N}\frac{dw_{1N}}{l_{w }}\nn
A'_{2N} &=& -A_{v}(y_2)d\psi_N+A_{v} D\psi_N+l_v\lambda_{2N}\frac{dw_{2N}}{l_{w }}\nn
A'_{1S} &=& -A_{v}(y_1)d\psi_S+A_{v} D\psi_S+l_v\lambda_{1S}\frac{dw_{1S}}{l_{w }}\nn
A'_{2S} &=& -A_{v}(y_2)d\psi_S+A_{v} D\psi_S+l_v\lambda_{2S}\frac{dw_{2S}}{l_{w }}~.
\eea
Here $\lambda_{1N}, \lambda_{2N}, \lambda_{1S}, \lambda_{2S}$ are constants to be fixed
by the requirement that the $(1/l_v)A'$ patch together to give a connection one-form. We choose
$\lambda_{1N}=\lambda_{2N}=\lambda_{1S}=\lambda_{2S}\equiv \lambda$ with
\bea\label{salt}
\frac{A_{v}\left(y_{1}\right)-A_{v}\left(y_{2}\right)}{l_{v}}+\lambda q & =-a\nn
\frac{2A_{v}\left(y_{2}\right)}{l_{v}}+\lambda p & =b.\eea
%
where $a,b$ are integers satisfying $ap+bq=1$, which is possible because
of \reef{pdtrick}.
Consider first the overlap of $U_{1N}$ with $U_{2N}$. On this overlap we have
\bea\label{jh}
\frac{1}{l_{v}}\left[A'_{2N}-A'_{1N}\right]= -a d\psi_N~.
\eea
Since $\psi_N$ has period $2\pi$ and $a$ is an integer, we see that
the two connections do indeed differ by a $U(1)$ gauge transformation.
Next consider the overlap of $U_{2S}$ with $U_{2N}$. Here we have
\bea\label{jh2}
\frac{1}{l_{v}}\left[A'_{2S}-A'_{2N}\right] =-bd\phi~.
\eea
It is illuminating to compare with equations (\ref{abeqn}) and
(\ref{atorsion}), (\ref{btorsion}) in appendix \ref{topy}.
In particular, we see that \reef{jh} and \reef{jh2}
give\footnote{For a more detailed explanation of the relation
between the transition functions (\ref{jh}), (\ref{jh2})
and torsion Chern classes, we refer to appendix \ref{lensy}.} the torsion Chern classes over $E_f$ and $E_2$, respectively.
As a further check, we compute
\bea\label{1overlap}
\frac{1}{l_{v}}\left[A'_{1S}-A'_{1N}\right] &=& -(b-2a)d\phi~,
\eea
which is equivalent to the Chern number of the $w$-fibration
over $\Sigma_1$ being $p+2q$ and agrees with (\ref{abtorsion}).
Note that, conversely, if one allows general $\lambda$ in (\ref{lambdas})
and instead imposes that the connections differ by $U(1)$ gauge
transformations (\ref{jh}), (\ref{jh2}) on the overlaps, then
one finds the solution (\ref{salt}).

Now consider $\hat{M_6}$, where we divide the period of $v$ by $q(p+q)$.
Note immediately that the connection form on $U_{2N}\cap U_{1N}$ is
\bea
\frac{q(p+q)}{l_{v}}\left[A'_{2N}-A'_{1N}\right] = -a(p+q)\left[\frac{dw_{2N}}{l_{w}}-\frac{dw_{1N}}{l_{w}}\right]~.
\eea
Thus we may define
\bea
\frac{q(p+q)}{l_{v}}\hat{A}'_{1N} &=&\frac{q(p+q)}{l_v}A'_{1N} + a(p+q)\frac{d w_{1N}}{l_{w}}\nn
\frac{q(p+q)}{l_v}\hat{A}'_{2N} &=&\frac{q(p+q)}{l_v}A'_{2N} + a(p+q)\frac{d w_{2N}}{l_{w}}~.
\eea
These are good gauge transformations on each patch. We see that $\hat{A}'_{1N}$ and $\hat{A}'_{2N}$ agree on the overlap,
and thus define a \emph{globally} defined one-form on
the complement of $\{\theta=\pi\}$. In particular, this shows
explicitly that the $\hat L$ circle bundle over $E_f$, i.e. the $v$ bundle over $E_f$ (with the period
above), is trivial\footnote{Note that we only need to quotient the period of
$v$ by $q$ to be able to do this, not $q(p+q)$, as expected from the comment
in footnote 8.}.
A globally defined connection one-form
is provided by $\tfrac{q(p+q)}{l_v}\hat{A}'$ above, restricted to $\{\theta=0\}$.

Remarkably, the factors of $a$ and $b$
in $\hat{A}'$ now cancel, and the connection form
reduces to
\bea
\hat{A}'_{1N}= -A_{v}(y_1)d\psi_N+A_{v}D\psi_N+\frac{l_v}{2q(p+q)}\frac{dw_{1N}}{l_{w}}~.
\eea
We are now in a position to calculate the integral of the three-form flux over $V_2$. 
Recall that the submanifold $V_2$ is obtained as a section
of the $\hat L$ circle bundle over $F_2\equiv E_f$.
In $F^{(3)}$ we therefore 
set $\theta=0$ and replace $Dv=dv-A'$ with $-\hat{A}'_{1N}$. After some calculation
we obtain
\bea\label{rone}
\frac{1}{(2\pi l_s)^2 g_s}\int_{V_2} F^{(3)} = -\frac{L^2}{l_s^2g_s}\frac{ (p+q)^2}{pq(p+2q)} \equiv -\frac{M_2}{q}~.
\eea
where $M_2$ is a positive integer.

It remains to calculate the integral over the submanifold $V_1$,
obtained as a section of the $\hat L$ circle bundle over the submanifold $F_1$ obtained by smoothing out 
$E_2\cup-E_f$.
We cover $V_1$ by 3 patches: $U_{1N}$, $U_{2N}$ and $U_{2S}$.
These will cover the northern hemisphere, equatorial strip, and southern hemisphere patches, respectively,
of the $S^2$ we get by gluing $\Sigma_2$ to $-\Sigma_f$.
This is illustrated in Figure 1.
To be more precise we will cover most of
$\Sigma_f$ in $U_{1N}$ by setting $\theta=0$,
letting $y\in[y_1,y_2-\epsilon]$ with $\psi_N$
the azimuthal angle. We will cover most of $\Sigma_2$ in $U_{2S}$ by setting $y=y_2$, letting
$\theta\in [\delta,\pi]$ with $\phi$ the azimuthal angle. Here
$\epsilon,\delta>0$ are small. On the overlap in $U_{2N}$
the equatorial strip, $Eq$,
is the line in the $\delta,y$ plane stretching
from $(\theta,y)=(\delta,y_2)$
to $(\theta,y)=(0,y_2-\epsilon)$, over which there is an azimuthal angle -- at the first end of this line
it is $\phi$ and at the other end it is $\psi_N$.
In fact, on this strip the azimuthal angles get identified via
\bea
\phi=-\psi_N~,
\eea
with the sign corresponding to an orientation flip.

\begin{figure}[!ht]
\begin{center}
\psfig{file=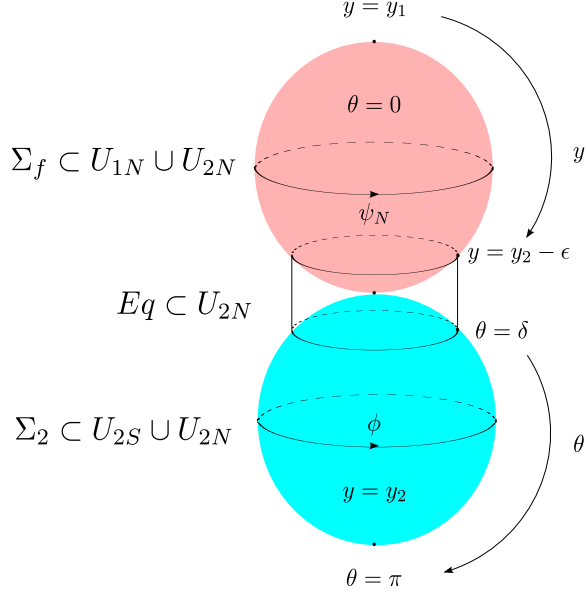,width=3in}
\end{center}
\caption{Desingularisation of $\Sigma_2\cup -\Sigma_f$.}
\label{fig1}
\end{figure}

We first examine the overlaps
\bea
\frac{q(p+q)}{l_v}\left[A'_{2N}-A'_{1N}\right] &=& -aq(p+q)d\psi_N = q((b-a)q-1) d\psi_N\nn
\frac{q(p+q)}{l_v}\left[A'_{2S}-A'_{2N}\right] &=& -bq(p+q)d\phi = -q((b-a)p+1)d\phi~.
\eea
This leads us to define
\bea
\frac{q(p+q)}{l_v}\tilde{A}'_{2N} = \frac{q(p+q)}{l_v}{A}'_{2N} + qd\psi_N - q(b-a)\frac{dw_{2N}}{l_{w}}~,
\eea
which is obtained via a good gauge transformation on this patch.
We then find
\bea
\frac{q(p+q)}{l_v}\left[\tilde{A}'_{2N}-A'_{1N}\right] &=& q^2(b-a)d\psi_N - q(b-a)\frac{dw_{2N}}{l_{w}} \nonumber \nn
&=& -q(b-a)\frac{dw_{1N}}{l_{w}}\\
\frac{q(p+q)}{l_v}\left[A'_{2S}-\tilde{A}'_{2N}\right] &=& -pq(b-a)d\phi + q(b-a)\frac{dw_{2N}}{l_{w}} \nonumber \\ &=&
q(b-a)\frac{dw_{2S}}{l_{w}}~.
\eea
This prompts us to define
\bea
\frac{q(p+q)}{l_v}\tilde{A}'_{1N}& = &\frac{q(p+q)}{l_v}{A}'_{1N} -q(b-a)\frac{dw_{1N}}{l_{w}}\nn
\frac{q(p+q)}{l_v}\tilde{A}'_{2S}& = &\frac{q(p+q)}{l_v}{A}'_{2S} -q(b-a)\frac{dw_{2S}}{l_{w}}~,
\eea
which are again obtained via good gauge transformations on the patches.
After all this, $\tilde{A}'$ is a globally defined one-form on $F_1$, and thus we see
explicitly that the $v$ bundle trivialises over it
since\footnote{Note that to obtain this result we only needed to quotient the period of $v$ by $p+q$ here, not $(p+q)q$.
In particular, all of the above gauge transformations are well-defined.}
we have divided the period by $(p+q)q$.
Moreover, one finds that $a$ and $b$ end up completely
cancelling, and that the correct connection form to use on $U_{1N}$ and $U_{2S}$ is
\bea
\tilde{A}'_{1N}& = & -A_{v}(y_1)d\psi_N+A_{v}D\psi_N-\frac{l_v}{2q(p+q)}\frac{dw_{1N}}{l_{w}}\nn
\tilde{A}'_{2S}&=& -A_{v}(y_2)d\psi_S+A_{v}D\psi_S-\frac{l_v}{2q(p+q)}\frac{dw_{1S}}{l_{w}}~.
\eea
By taking $\epsilon,\delta\to 0$
we effectively use the gauge $\tilde{A}'_{1N}$ over $E_f$ and $\tilde{A}'_{2S}$ over $E_2$ and then consider the
result for $E_2$ minus the result for $E_f$.
After some calculation this gives the period
\bea\label{rtwo}
\frac{1}{(2\pi l_s)^2 g_s}\int_{V_1} F^{(3)} = -\frac{L^2}{l_s^2g_s}\frac{(p+q)^3}{pq^2(p+2q)}      \equiv -\frac{M_1}{p+q}~,
\eea
where $M_1$ is a positive integer.

Consistency of \reef{rone} and \reef{rtwo}
implies that we choose
\bea \label{consflux}
M_1=M(p+q)^2,\qquad M_2=Mq^2
\eea
for some positive integer $M$ and
\be
\frac{{L}^2}{l_s^2 g_s}=\frac{pq^2(p+2q)M}{(p+q)^2}~.
\ee
We have thus recovered the results \reef{resone} and \reef{restwo},
which is very satisfying.

\subsection{Taking the limit $Q\to 0$}
We have noted that the parameter $Q\ne 0$ corresponds to an exactly marginal deformation of the
underlying dual $(0,2)$ SCFT. In particular, the solutions, with the internal manifold having topology
$S^3\times S^3\times S^1$, and all other parameters fixed, are smoothly connected with each other as $Q$ varies
and they all have the same central charge.
It is interesting to observe that the value of the central charge is precisely the same as for the solutions with $Q=0$
that were analysed in \cite{ajn}. It is therefore tempting to conclude that the solutions with $Q\ne 0$ correspond to exactly marginal
deformations of the SCFTS dual to the $Q=0$ solutions. While this may in fact be the correct
interpretation, it seems difficult to draw this conclusion based on the results of this paper.
In particular, taking the limit $Q\to 0$ does not simply lead to the $Q=0$ solutions discussed in \cite{ajn}, as we now explain. 

It is certainly the case that setting $Q$ to zero in the local solutions given by \reef{sol1}, \reef{startingpoint}, \reef{dial}
and \reef{threeformtdual} 
one obtains the
local solutions with $Q=0$ studied in \cite{ajn}. However, we also need to check what happens to the global
identifications we have made on the coordinates, and also the quantisation of the flux. 
Recall that in the solutions of \cite{ajn} with $Q=0$ 
the internal manifold has topology $S^2\times S^3\times T^2$, with the $S^2\times S^3$ factor realised in the same
way as the $M_5$ factor for the $Q\ne 0$ solutions studied here. It is simple to check that taking the limit
$Q\to 0$ in the spaces $B_4$ and $M_5$, labeled by $p,q$, studied in sections 2.2 and 2.3, respectively, leads to the corresponding spaces in the $Q=0$ solutions of \cite{ajn}. 
The problem arises in taking the limit of the $v$ circle fibration. 
In \reef{pdtrick} we impose\footnote{If one instead imposes that this has twist 0 then \reef{pdtrick}, \reef{hull} and \reef{mancity}
imply that $q=0$ or $p=-q$. Returning to \reef{fperiods}
we see that imposing $q=0$ and using \reef{hull} would imply $\beta=0$ 
which is excluded. Similarly imposing $p=-q$ is not compatible with \reef{hull}. }
that this has twist 1 over the base $M_5 = S^2 \times S^3$. This is impossible as $Q\to 0$, because the
connection $A_v\to 0$ and correspondingly, the period $l_v$ goes to zero in \reef{manu}. From
\reef{convchange} we therefore conclude that the period of the $u^1$ circle goes to zero in this limit and we do not match onto
the $Q=0$ solutions of \cite{ajn} where the $u^1$ circle in the $T^2$ factor had finite period.   

It is worth pointing out that in the $Q=0$ solutions of \cite{ajn}, with topology $S^2\times S^3\times T^2$, only the volume of the $T^2$ 
was fixed by the flux quantisation. The shape of the $T^2$ therefore corresponds to exactly marginal deformations.
It might be the case that going to the limit where the period of the $u^1$ circle goes to zero is connected to
the $Q\to 0$ limiting solutions. However, in this limit the supergravity approximation is clearly breaking down and one needs
to analyse string corrections before any definite conclusions can be drawn. It would be interesting to study this further.

\section{More general identifications}
In this section we will generalise the class of solutions
that we have already constructed. We return to the local
solution \reef{metnewcoords}, \reef{dial}, \reef{tfnewcoords}
and then employ the general linear coordinate transformation
\bea
w= & h\, w'+r\frac{Q}{Z}\, v'\nn
v= & s\, w'+\frac{t}{2\beta}\, v'
\eea
for constant $r,t,s,h$ with
\bea\label{dnz}
\Delta=  h\frac{t}{2\beta}-r\frac{Q}{Z}s\neq0~.
\eea
The idea is to now make appropriate periodic identifications
of the new coordinates $v',w'$.
As we shall see this will embed our solutions of type IIB string string theory
of the last two sections into larger families.

We first observe that
\bea
Dw&= & h\, Dw'+r\frac{Q}{Z}\, Dv'\nn
Dv&= & s\, Dw'+\frac{t}{2\beta}\, Dv'\eea
where we have defined
\bea
Dw'&= & dw'-A_{w'}\, D\psi\nn
Dv'&= & dv'-A_{v'}\, D\psi
\eea
with
\bea
A_{w'}&= & \frac{t}{2\beta\Delta}\, A_{w}-r\frac{Q}{Z\Delta}\, A_{v}\nn
A_{v'}&= & \frac{h}{\Delta}\, A_{v}-\frac{s}{\Delta}\, A_{w}~.\eea
We now construct $M_5$ as a circle fibration, with circle parametrised by $w'$,
over $B_4$ and then construct $M_6$ as a circle fibration, with circle
parametrised by $v'$, over $M_5$. It is straightforward to write the
metric in the primed coordinates and then appropriately ``complete the square''
to make this fibration structure manifest in the metric.
However, we will not need the
explicit details. Observe that what will become the
globally defined angular one-form on $M_5$ for the $w'$ circle fibration is
$Dw'$.
After completing the square in the metric on $M_6$ we obtain an expression for what will become the globally defined angular one-form corresponding to the $v'$ circle fibration and it has the form
\be
dv-A_{v'}D\psi-k(y)Dw'
\ee
for some smooth function $k(y)$ that can easily be determined.
The connection one-form on $M_5$ for this circle fibration
is thus $A_{v'}D\psi+k(y)Dw'$. This will turn out to be
a local connection one-form on the same circle bundle
as that for the connection one-form $A_{v'}D\psi$, since
$kDw'$ will be globally defined on $M_5$ (in particular, the corresponding
curvature two-forms are in the same cohomology class on $M_5$). Below, for
convenience, we will use the connection one-form $A_{v'}D\psi$.

The analysis now proceeds in an almost identical fashion as in the last sections, so we can be brief. We
choose the period of the $w^{\prime}$ circle to be $2\pi l_{w'}$ so that
$l_{w'}^{-1}A_{w'}D\psi$ is a connection on a $U(1)$ fibration.
We demand that $(2\pi l_{w})^{-1}d(A_{w}\, D\psi)$ has integer
periods on $B_4$, as in \reef{fperiods}, with primes on all $w$,
for some integers $p,q$, now {\it not necessarily positive}.
When $r+t=0$ we have $q=0$, while when $r-t=0$ we have $p+q=0$
and these cases
require a separate analysis which we will return to later.
We thus continue here with $r\ne \pm t$ and conclude
that\footnote{Note that if we choose $t=\beta(Z-2)/(Z-1)$, $r=-\beta Z/(Z-1)$, $h=(Z-2)/2(Z-1)$
and $s=Z(Z-2)/4Q\beta (Z-1)$, then we have $w'=z$, $v'=u^1$, where $z, u^1$ are the coordinates
that we started with in \reef{startingpoint}. In this case equation \reef{betaexp} becomes
$\beta=(1-Z)/(1+X)$ and $l_{w'}=2(1+X)/q(X+Z)(2+X-Z)$,
where $X=p/q$ and this agrees with the results in equation (4.22) of \cite{ajn}.}
\bea\label{betaexp}
\beta &=&\frac{t-r}{t+r}\frac{q}{p+q}\nn
l_{w'}&=&\frac{(2-Z)(r+t)}{2q(1-Z)(1-\beta^2)\Delta}
\eea
The topology of $M_5$ is again $S^3\times S^2$.
For the generator of $H^{3}\left(M_{5},\mathbb{Z}\right)$ we can use the primed version of
\reef{m5threegen}.

We now turn to the $v'$ circle fibration over $M_5$ to give $M_6$.
We let $v'$ be a periodic coordinate with period $2\pi l_{v'}$, and
the connection one-form is given by $l_{v'}^{-1}A_{v'} D\psi$.
To ensure that the circle fibration is well-defined and
that $M_6=S^3\times S^3$ we impose the primed version of
\reef{pdtrick} to conclude that
\be
l_{v'} =\frac{2qQ}{\left(r+t\right)\left(1-Z\right)}.
\ee

Now we determine the flux quantisation conditions. The electric flux quantisation condition
\reef{fqnc} fixes the period of $u^2$ as before:
\be\label{el}
n_1=\frac{L^6}{l_s^6 g_s}\frac{Q}{8\pi^2\beta(1-\beta^2)^2}\Delta u^2~.
\ee
For the magnetic flux quantisation, we follow the same procedure as before,
by introducing explicit coordinate patches and considering integrals on submanifolds
of $\hat M_6=M_6/\bbZ_{(p+q)q}$. By following the same steps as in section 3.2 we find
that
\bea
\frac{1}{\left(2\pi l_{s}\right)^{2}g_{s}}
\int_{V_{2}}F^{(3)}&=&\frac{L^{2}}{l_{s}^{2}g_{s}}\frac{1}{q}\frac{1}{\beta^{2}-1}\equiv -\frac{M_2}{q}\nn
\frac{1}{\left(2\pi l_{s}\right)^{2}g_{s}}\int_{V_{1}}F^{\left(3\right)}&=&\frac{L^{2}}{l_{s}^{2}g_{s}}\frac{1}{q}\frac{r+t}{r-t}\frac{1}{\beta\left(1-\beta^{2}\right)}\equiv -\frac{M_1}{p+q}\eea
for integers $M_2,M_1$.
Consistency implies that we must have
\bea\label{rcon}
\frac{M_2}{M_1}=\beta^2=\frac{ (r-t)^2}{(r+t)^2}\frac{q^2}{(p+q)^2},
\eea
which implies that $(r+t)^2/(r-t)^2$ must be rational,
and that the length scale is fixed by
\be\label{lengths}
\frac{L^2}{l_s^2 g_s}=(1-\beta^2)M_2=\frac{(M_1-M_2)M_2}{M_1}
\ee

The central charge can now be calculated, and we find that it can be expressed
as
\be\label{cench}
c=6n_1\frac{(M_1-M_2)M_2}{M_1}~.
\ee
In particular we note that, in addition to $Q$, there is also no dependence on the parameters $r,s,t$
and $h$. We note that the only restrictions on these parameters
is \reef{dnz}, \reef{rcon} with $\beta$ given in
\reef{betaexp} satisfying $0<\beta<1$.
We have thus constructed large continuous
families of solutions that are dual to SCFTs. Note that, in general, the
solutions of this section are not exactly marginal deformations of those
in section 3: for example, in section 3 we saw that the magnetic three-form
flux quantum numbers were constrained to be of the form \reef{consflux}, whereas
here there is no such constraint.

\subsection*{When $r=-t$:}
When $r=-t$, we have $A_{w'}(y_1)=A_{w'}(y_2)$ and hence
in considering the $w'$ circle fibration over $B_4$ to construct
$M_5$ we find that the period over $C_2=\Sigma_f$ vanishes, $q=0$.
We choose $p=1$, so that the period over $C_1=\Sigma_2-\Sigma_f$ is one,
and hence $M_5=S^3\times S^2$, which implies that
\be
l_{w'}=2 A_{w'}(y_2)=-\frac{t(Z-2)}{\beta(\beta^2-1)(Z-1)\Delta }~.
\ee
At this stage, there is no restriction on the parameter $\beta$
(apart from the usual $0<\beta<1$).
We now find on $M_5$ that $E_1, E_2\cong S^3$, and $[E_1]=[E_2]$
generate $H_3(M_5,\bbZ)$. On the other hand, now $E_f\cong S^1\times S^2$
(and hence there is a section of the $w'$ circle fibration over $\Sigma_f$).
The generator of $H^2(M_5,\bbZ)$ is $\tau=\sigma_f$ {\it i.e.}
$a=1, b=0$, in the notation of section 2.3.

In order to construct $M_6=S^3\times S^3$, we can again fix the period
of the $v'$ circle using $\Phi$ as in \reef{pdtrick} and we find that
\be
l_{v'}=A_{v'}(y_2)-A_{v'}(y_1)=-\frac{\beta Q}{(Z-1)t}~.
\ee
It is now easier to find representatives of the two generators of
$H_3(M_6,\bbZ)$,
and we won't have to consider a quotient of $M_6$ in order
to impose the flux quantisation conditions.
In particular, one generator of $H_3(M_6,\bbZ)$, $W_1$,
can be taken to be, as above,
the section of the $v'$ circle fibration over a desingularised version
of $E_2\cup -E_f$. For the other generator, $W_2$,
we can take the $v'$ circle bundle
over the section $s(S)$ on $M_5$ where\footnote{Before, when $[S]=q[\Sigma_2]+
p[\Sigma_f]$, it was not clear how to take a smooth representative for $S$.}
$S=\Sigma_f$. We note that two other three-cycles $W'$, $W''$ are obtained by
considering a section of the $v'$ circle fibration over $E_1$, $E_2$, respectively:
we shall show that $[W']=[W'']=[W_1]+[W_2]$.

We now introduce patches in exactly the same way as section 3.2. The analogue
of \reef{qtwistnew} now reads $w'_{2N}=w'_{1N}$ and
we explicitly see that the $w'$ circle fibration is indeed trivial over
$\Sigma_f$. To obtain the section $s[\Sigma_f]$ we can simply set
$w'_{2N}=constant$.

Moving to $M_6$, we have the analogue of the connection one-forms as
in \reef{lambdas}, \reef{salt} with $a=1$, $b=0$, $p=1$, $q=0$.
Equation \reef{jh2}
shows that the $v'$ circle
fibration is indeed trivial over $E_2$ and we can take a section to
obtain the three-cycle $W''$. One can then obtain the integral of the
three-form flux over $W''$ by using the connection one form
$A^\prime_{2N}$, and after a calculation we find
\begin{equation}\label{dub}
\frac{1}{\left(2\pi l_{s}\right)^{2}g_{s}}\int_{W''} F^{\left(3\right)}
=-\frac{L^{2}}{l_{s}^{2}g_{s}}
\frac{1}{\beta^2}~.
\end{equation}
The $v'$ circle fibration is also trivial over $E_1$. Indeed,
after considering \reef{jh2} we can see that the connection one-form
\be
A'_{1S}+2l_{v'}\frac{dw'_{1s}}{l_{w'}}
\ee
is a globally defined one-form on $E_1$. We can use this gauge to calculate the
integral over $W'$ and we find exactly the same result as for $W''$.

To calculate the integral of the flux over the three-cycle $W_2$,
the $v'$ circle bundle over the section $s(\Sigma_f)$, we just need to
set $w'_{2N}=constant$ in the expression for the three-form and then integrate.
We therefore impose
\begin{equation}\label{dub2}
\frac{1}{\left(2\pi l_{s}\right)^{2}g_{s}}\int_{W_2} F^{\left(3\right)}
=\frac{L^{2}}{l_{s}^{2}g_{s}}
\frac{1}{\left(1-\beta^{2}\right)}=M_2~.
\end{equation}

To carry out the flux integral over $W_1$, a section of the
$v'$ circle fibration over $E_2\cup -E_f$, we define
\bea
\frac{1}{l_{v'}}\tilde{A}_{2N}^{\prime} & =&\frac{1}{l_{v'}}A_{2N}^{\prime}+d\psi_N+\frac{dw_{2N}}{l_w}\nn
\frac{1}{l_{v'}}\tilde{A}_{1N}^{\prime} & =&\frac{1}{l_{v'}}A_{1N}^{\prime}+\frac{dw_{1N}}{l_w}\nn
\frac{1}{l_{v'}}\tilde{A}_{2S}^{\prime} & =&\frac{1}{l_{v'}}A_{2S}^{\prime}+\frac{dw_{2S}}{l_w}~.
\eea
Then $\tilde A^{\prime}$ is a global one-form on $E_2\cup -E_f$.
To calculate the integral of flux over the section over the $v'$ circle bundle over $E_2\cup -E_f$ we use $\tilde{A}_{2S}^{\prime}$ on $E_2$ and
$\tilde{A}_{1N}^{\prime}$ on $E_f$. We find
\begin{equation}\label{dub1}
\frac{1}{\left(2\pi l_{s}\right)^{2}g_{s}}\int_{W_{1}}F^{\left(3\right)}
=-\frac{L^{2}}{l_{s}^{2}g_{s}}
\frac{1}{\beta^2\left(1-\beta^{2}\right)}=-M_1~.
\end{equation}
Comparing \reef{dub}
with \reef{dub2} and \reef{dub1}, we can deduce the
homology relation $[W'']=[W']=[W_1]+[W_2]$, as mentioned above.

Consistency of \reef{dub2} and \reef{dub1} implies that the length
scale of the solution is again as in \reef{lengths} and that $\beta^2$
is rational
\be
\beta^2=\frac{M_2}{M_1}~.
\ee
The electric flux quantisation condition is given again by \reef{el} and
the central charge takes the form \reef{cench}.

\subsection*{When $r=t$:}

When $r=t$, we have $A_{w'}(y_1)=-A_{w'}(y_2)$ and hence
in considering the $w'$ circle fibration over $B_4$ to construct
$M_5$ we find that the period over $C_1=\Sigma_2-\Sigma_f$ vanishes,
$p+q=0$. We choose $q=1$ so that the period over $C_2=\Sigma_f$ is one,
and hence $M_5=S^3\times S^2$, which implies that
\be
l_{w'}=\frac{-t(Z-2)}{(\beta^2-1)(Z-1)\Delta}
\ee
with no restriction on the parameter $\beta$.
We now find $E_1, E_2, E_f\cong S^3$, and $-[E_1]=[E_2]=[E_f]$
generate $H_3(M_5,\bbZ)$. The generator of $H^2(M_5,\bbZ)$ is
$\tau=b\sigma_2+a\sigma_f$ with $b-a=1$.

In order to construct $M_6=S^3\times S^3$, we find that
the period of the $v'$ circle is
\be
l_{v'}=A_{v'}(y_1)+A_{v'}(y_2)=-\frac{Q}{(Z-1)t}~.
\ee
For the generators of $H_3(M_6,\bbZ)$ we can take
$W_1$ to be the $v'$ circle fibration over the a representative of
the section $s(S)$ of $M_5$, with $[S]=[\Sigma_2]-[\Sigma_f]$.
For $W_2$ we take a section of the $v'$ circle fibration
over $E_f$. We note that we can also obtain three-cycles $W'$, $W''$ which
are obtained by considering sections of the $v'$ circle fibration over $E_1$, $E_2$ respectively:
we shall see that $-[W']=[W'']=[W_1]+[W_2]$.

We again introduce patches in exactly the same way as section 3.2.
The connection one-forms are as in \reef{lambdas}, \reef{salt}
with $b-a=1$ and $q=-p=1$. By taking $a=0$, $b=1$, we see from \reef{jh}
that $A'_{2N}$  is a globally defined connection one-form on
$E_f$. Calculating the flux integral we find
that we should impose
\begin{equation}\label{dub2t}
\frac{1}{\left(2\pi l_{s}\right)^{2}g_{s}}\int_{W_2} F^{\left(3\right)}
=-\frac{L^{2}}{l_{s}^{2}g_{s}}
\frac{1}{\left(1-\beta^{2}\right)}=-M_2~.
\end{equation}
To integrate the flux integrals for $W'$ one should take $b=2$, $a=1$ while for
$W''$ we should take $b=0$, $a=-1$ and we find
\begin{equation}
-\frac{1}{\left(2\pi l_{s}\right)^{2}g_{s}}\int_{W'} F^{\left(3\right)}
=\frac{1}{\left(2\pi l_{s}\right)^{2}g_{s}}\int_{W''} F^{\left(3\right)}
=\frac{L^{2}}{l_{s}^{2}g_{s}}
\frac{1}{\beta^{2}}~.
\end{equation}

We now turn to the flux integral over $W_1$. For $S$ we desingularise
$\Sigma_2-\Sigma_f$ as in Figure 1. By making the gauge transformation
$w'_{2N}\to w'_{2N}-l_{w'}d\psi_N$ in \reef{dubs}, we find that we obtain a globally defined
connection one-form on $S\subset M_5$ and hence we can take a section. $W_1$
is obtained by considering the $v'$ circle fibration over this section. Thus to
calculate the flux integral, one should set $w'_{1N}=constant$ in $Dw'_{1N}$
for the $\Sigma_2$ piece and $w'_{1S}=constant$ in $Dw'_{1S}$
for the $\Sigma_f$ piece. After doing this we find
\begin{equation}\label{dub1t}
\frac{1}{\left(2\pi l_{s}\right)^{2}g_{s}}\int_{W_{1}}F^{\left(3\right)}
=-\frac{L^{2}}{l_{s}^{2}g_{s}}
\frac{1}{\beta^2\left(1-\beta^{2}\right)}=-M_1~.
\end{equation}
We thus find the same conditions as for the $r=-t$ case above.

\section{Final Comments}

We have analysed in detail some local
supersymmetric $AdS_3$ solutions of type IIB supergravity, first found
in \cite{ajn}, that have non-vanishing dilaton and $RR$ three-form flux.
We have shown that the parameters can be chosen and coordinates identified in
such a way that the solutions extend to give rich classes of globally defined solutions
of the form $AdS_3\times_w( S^3\times S^3\times S^1)$
with properly quantised flux. We have shown that the solutions depend
on continuous parameters and are hence dual to continuous families of SCFTs
in two spacetime dimensions with $(0,2)$ supersymmetry.

Although the internal compact spaces are diffeomorphic
to $S^3\times S^3\times S^1$, the diffeomorphisms are far from apparent in the local
coordinates that the solutions are presented in. It seems unlikely to us that there is a
simple change of coordinates that will make the topology more manifest.
In this paper we used a number of techniques to illuminate various aspects of the topology which, in particular,
allowed us to find a workable procedure to impose flux quantisation. It seems likely
that our approach, or generalisations thereof, will be very useful in other contexts.

In section 4 we considered identifications on the coordinates after we made a general linear
transformation on the $v,w$ coordinates. It is worth pointing out that we
could consider more general linear coordinate
transformations that also involve the $u^2$ coordinate. This
will lead to larger families of solutions that would be worth exploring. It seems possible that some
of these solutions can be obtained as $\beta$-deformations using the techniques of
\cite{lm}. In fact returning to the solutions in section 2 and 3,
one might wonder if $Q$ corresponds to a $\beta$-deformation.
One way to see that it is not is to return to the local solutions as written down at the beginning
of section 4 of \cite{ajn},
which are obtained after two T-dualities on the solutions we have discussed in this paper.
In this duality frame only the metric and the self-dual five-form are non-trivial for any $Q$,
and in particular the dilaton is constant.
However, looking at equation (A.16) of \cite{lm} we see that the $\beta$-deformation
activates a  non-trivial dilaton and three-form.

It is an important outstanding issue to identify the dual $(0,2)$ SCFTs for
the solutions discussed here and in \cite{Gauntlett:2006af,Gauntlett:2006qw,Gauntlett:2006ns,ajn}.
In the duality frame that we have used in this paper, the amount of supersymmetry that
is preserved combined with the fluxes that are active suggests that the dual SCFTs might
arise on a D1-D5-brane system that is wrapped on a holomorphic four-cycle in a Calabi-Yau four-fold.
While we remain hopeful that progress
will be made in this direction, we note that the SCFTs dual to the
much simpler type IIB $AdS_3\times S^3\times S^3\times S^1$ solutions of
\cite{Cowdall:1998bu}, which have $(4,4)$ supersymmetry, are still
not well-understood, despite interesting progress
\cite{Boonstra:1998yu,de Boer:1999rh,Gukov:2004ym,Berg:2006ng}.

The $AdS_3$ solutions with $Q=0$, that were analysed in \cite{ajn}, and with $Q\ne 0$ that we have discussed here,
can be generalised further and we have presented some details in appendix C. It will be interesting to carry out a complete
analysis of the conditions for regularity and flux quantisation conditions for these more general solutions.

\subsection*{Acknowledgements}
We would like to thank Jaume Gomis,
Dominic Joyce, Spiro Karigiannis, Nakwoo Kim, Tommaso Pacini and David Tong for helpful
discussions, and Bob McNees for help with Inkscape. We would also like to
thank one of the referees for insightful comments which led to the discussion
in section 3.3.
AD would also like to thank the Institute for Mathematical Sciences at
Imperial College for hospitality. JPG would like to thank the Perimeter
Institute for hospitality.
JPG is supported by an EPSRC Senior Fellowship and a
Royal Society Wolfson Award. JFS is supported by a Royal
Society University Research Fellowship.

\appendix

\section{$U(1)$ bundles over Lens spaces}\label{lensy}
In this section we briefly review the Lens spaces $S^3/\bbZ_q$,
which appear throughout the main text, and also the construction
of $U(1)$ principal bundles over these manifolds.

We construct $S^3/\bbZ_q$ as the total space of a $U(1)$ bundle over $S^2$ with
Chern number $q$. Let $\theta$, $\phi$ be standard coordinates on $S^2$, and cover
the $S^2$ with two patches: $V_N$ which excludes the south pole $\theta=\pi$, and $V_S$ which excludes the north pole $\theta=0$.
We then consider the products $S^1\times V_N$, $S^1\times V_S$, and
on each space define the one-forms
\bea
D\nu_N &=& d\nu_N - \frac{q}{2}(1-\cos\theta)d\phi\nn
D\nu_S &=& d\nu_S + \frac{q}{2}(1+\cos\theta)d\phi~.
\eea
Here $\nu_N$ and $\nu_S$ are coordinates on the $S^1s$, each with period $2\pi$. If we now glue the
two patches together via
\bea
\nu_S-\nu_N = -q\phi
\eea
on the overlap then note that
\bea
D\nu = D\nu_N = D\nu_S
\eea
extends to a global one-form on the whole manifold, because the two one-forms agree on the overlap. This is a global
connection form on the total space of the $U(1)$ principal
bundle $\mathrm{p}: S^3/\bbZ_q\rightarrow S^2$ with $U(1)$ fibre parametrised by $\nu$, and is sometimes
also called the global angular form.

Now consider the connection form
\bea\label{connect}
A_N &=& \frac{a}{2}(1-\cos\theta)d\phi\nn
A_S &=& -\frac{a}{2}(1+\cos\theta)d\phi
\eea
on the base $S^2$.
This has Chern number $a\in \bbZ$ over the base $S^2$. We denote
the corresponding $U(1)$ principal bundle by $P$. We may pull back $P$ to
a $U(1)$ bundle $\mathrm{p}^*P$ over $S^3/\bbZ_q$.
Pulling back the connection (\ref{connect}), on the overlap
one finds
\bea\label{Lenscon}
A_S - A_N = -a d\phi = \frac{a}{q}d(\nu_S - \nu_N)~.
\eea
Note that $a\nu_S/q$ is a multi-valued $U(1)$ function on the patch $S^1\times V_S$ unless
$a/q\in \bbZ$. If $a/q\in \bbZ$ then in each patch we can define
the new connection one-forms $A_S-ad\nu_S/q$ and $A_N-ad\nu_N/q$, and since they agree on the
overlap, this defines a globally defined connection one-form and hence $\mathrm{p}^*P$ is trivial.

Thus $\mathrm{p}^*P$ is trivial if and only
if $a\cong 0$ mod $q$.
One sees this in a more abstract way
by recalling that $U(1)$ principal bundles are classified
by $H^2(S^3/\bbZ_q,\bbZ)\cong H_1(S^3/\bbZ_q,\bbZ)\cong \bbZ_q$.
Thus $a\in \bbZ_q$ is precisely the Chern number of
$\mathrm{p}^*P$, and the latter bundle is torsion.
Because of this, the topology
cannot be measured
by integrating the curvature of a connection $A$ over a two-cycle --
to see torsion classes using the connection is more subtle.
This is explained in general in the paper \cite{freed}.
The latter reference implies that the torsion
first Chern class may be computed by picking a
\emph{flat} connection on $\mathrm{p}^*P$, and
then computing the  log of the holonomy
of this flat connection around the one-cycles that generate $H_1(S^3/\bbZ_q,\bbZ)$. We may shift to a flat connection here by defining
\bea
A_S^{\mathrm{flat}} &=& A_S +\frac{a}{q}D\nu_S = \frac{a}{q}d\nu_S\nn
A_N^{\mathrm{flat}} &=& A_N + \frac{a}{q}D\nu_N = \frac{a}{q}d\nu_N
\eea
Here we have added a global one-form $(a/q)D\nu$ to the
original connection -- we are simply picking a different connection on the  same bundle. Then $H_1(S^3/\bbZ_q,\bbZ)\cong \bbZ_q$ is generated by, for example, the $\psi_N$ circle
at $\theta=0$. Thus the log of the holonomy is
\bea
i\int_{S^1} A_{N}^{\mathrm{flat}} = \frac{2\pi i a}{q}\quad \mathrm{mod} \ 2\pi i~.
\eea
This implies that our connection above is $a$ times the generator of $\bbZ_q$.

Finally, we make a comment about quotients. First note that
quotienting the period of the $U(1)$ fibre coordinate
of $P$ by $q$ is the same as taking the $q$th
power of $P$.
In particular,
the $\bbZ_q$ quotient of the bundle $\mathrm{p}^*P$ over $S^3/\bbZ_q$
is then trivial. This follows simply because the connection on this bundle
in the two patches is $qA_S$ and $qA_N$, or after a gauge transformation
$qA_S-ad\nu_S$ and $qA_N-ad\nu_N$, and from \reef{Lenscon} we see that
this is a globally defined connection one-form, and hence the bundle is trivial.

\section{More on the topology of $M_5$}\label{topy}
Recall that, in the main text, $M_6$ is constructed as the total space of a
circle bundle $L$ over $M_5\cong S^3\times S^2$. Here $c_1(L)\in H^2(M_5,\bbZ)\cong
\bbZ$ is the generator, so that $M_6\cong S^3 \times S^3$. Although this
is straightforward as stated, the issue is that we have infinitely
many coordinate systems on $M_5$, labelled by the integers $p$ and $q$,
and the diffeomorphism $M_5\cong S^3\times S^2$ is not explicit for general $p$ and $q$.
For each $p$ and $q$ there are different naturally-defined
three-submanifolds of $M_5$ -- we are especially interested in three-submanifolds
since we would like to quantise the RR three-form flux.
In this appendix we consider these submanifolds in more detail, and in particular
determine the topology of $L$ restricted to them.

Consider restricting this circle bundle $L$ over
$M_5$ to one of the three-submanifolds of $M_5$: $E_1$, $E_2$ or $E_f$.
For example, take $E_f\cong S^3/\bbZ_q$. Recall this is itself a circle bundle over
$\Sigma_f\cong S^2$ with Chern class $q$.
There is an inclusion map $i_f:E_f\hookrightarrow M_5$, and
we can define a circle bundle $L_f$ over $E_f$ by pulling back
\bea
L_f \equiv i_f^* L~.
\eea
Since $E_f$ is a lens space, $E_f=S^3/\bbZ_q$,
circle bundles over $E_f$ are classified up to isomorphism by
\bea
c_1(L_f)\in H^2(E_f,\bbZ) \cong \bbZ_q~.
\eea
To compute this Chern class, recall that $c_1(L)=\pi^*\tau$,
where $\tau\in H^2(B_4,\bbZ)$ was defined in (\ref{lambdaeqn}).
Hence to compute $c_1(L_f)=i^*_f\pi^*(\tau)$ we may instead first restrict $\tau$ to
$\Sigma_f$, and then pull back using $\pi^*$ the corresponding circle bundle to $E_f$.
This is summarised by the following commutative square:
\bea
\begin{array}{ccc}
H^2(M_5,\bbZ) & \stackrel{i_f^*}{\longrightarrow} & H^2(E_f,\bbZ)\\
\pi^* \big\uparrow & & \big\uparrow \pi^*\\
H^2(B_4,\bbZ) & \stackrel{\iota_f^*}{\longrightarrow} & H^2(\Sigma_f,\bbZ)
\end{array}~.
\eea
Here we have denoted
the embedding of $\Sigma_f$ into $B_4$ by $\iota_f:\Sigma_f\rightarrow B_4$.
Then $\iota_f^*\tau$ defines an integer class in $H^2(\Sigma_f,\bbZ)\cong \bbZ$.
This in turn defines a circle bundle with Chern
number $a$, using (\ref{lambdaeqn}). Using the results in appendix A,
lifting this circle bundle to $E_f$ then gives a bundle with Chern number
\bea\label{atorsion}
a = c_1(L_f)\in H^2(E_f,\bbZ)\cong \bbZ_q~.
\eea
Thus the bundle $L$ restricted to $E_f$ is trivialisable only if $a=0$ mod $q$;
in other words, if $a=mq$ for some integer $m$. But if this were the case, then
we would have
\bea
(mp+b)q=1~.
\eea
This is only possible if $q=\pm 1$. Thus we see that for general $q$ it is not
possible to take a section of $L$ over $E_f$ to obtain a three-submanifold of
$M_6$.

One can do similar computations for the three-submanifolds
$E_1$ and $E_2$, with similar conclusions. We have
\bea
\label{abtorsion}
L_1\equiv i^*_1 L &,& \qquad c_1(L_1)=b-2a\in H^2(E_1,\bbZ)\cong \bbZ_{p+2q}\\
\label{btorsion}
L_2\equiv i^*_2 L &,& \qquad c_1(L_2)=b\qquad \ \in H^2(E_2,\bbZ)\cong \bbZ_p~.
\eea
Thus the corresponding bundles are trivial\footnote{This analysis assumes that $p$, $p+2q$, $q$
are non-zero.} if and only if $b=m_2p$, $b-2a=m_1(p+2q)$, respectively,
where $m_1,m_2\in \bbZ$,
which implies
\bea
p(a+qm_2)&=&1\nn
(p+2q)(a+m_1q)&=&1
\eea
respectively. These equations imply in particular that $p=\pm 1$ and $(p+2q)=\pm 1$.

We thus conclude that, for generic $p$ and $q$, the circle bundle $L$ restricted to
$E_1$, $E_2$ and $E_f$ is non-trivial, and thus we cannot globally take a section
of $L$. This means that these natural three-submanifolds of $M_5$ cannot be
used to construct natural three-submanifolds of $M_6$.

\section{More general $AdS_3$ solutions}

We first recall from \cite{Kim:2005ez}, \cite{ajn} the local data that is sufficient to construct supersymmetric $AdS_3$ solutions
of type IIB supergravity with non-vanishing five-form flux and complex three-form flux $G$.
We require a six-dimensional local K\"ahler metric $ds^2_6$ whose Ricci tensor satisfies\footnote{Changing the sign of the
last term leads to type IIB bubble solutions, as explained in \cite{ajn}. The construction in this appendix can be easily adapted to construct
bubble solutions.}
\be\label{miib}
\Box R-\frac{1}{2}R^{2}+{R}^{ij}{R}_{ij}+\frac{2}{3}G^{ijk}G_{ijk}^{\ast}=0
\ee
and $G$ must be a closed, primitive and $(1,2)$ three-form on the six-dimensional space.
We refer to \cite{Kim:2005ez}, \cite{ajn} for details
of how the full ten-dimensional solution is constructed from this data.

For the solutions that we have discussed in this paper, which we will now generalise,
the local six-dimensional Kahler metric has the form
\be\label{mbn}
ds^2_6=ds^2_4+ds^2(T^2)
\ee
where $ds^2(T^2)=(du^1)^2+(du^2)^2$ is the standard metric on a two-torus, $ds^2_4$ is a four-dimensional local K\"ahler metric,
and
\be\label{gbn}
G=d\bar u\wedge W
\ee
where $u=u^1+iu^2$ and $W$ is a closed, primitive $(1,1)$-form on the four-dimensional K\"ahler space.

Inspired\footnote{One can consider the scaling $\mu_3\to \epsilon \rho$, $q_3\to 1/\epsilon^2$, $\lambda\to\lambda/\epsilon^2$
in equation 5.10 of \cite{Gauntlett:2006ns} and then take $\epsilon\to 0$.}
by the six-dimensional K\"ahler metrics discussed in equation 5.10 of \cite{Gauntlett:2006ns},
we start with the ansatz for a four-dimensional K\"ahler metric given by
\begin{align}
ds_{4}^{2} & =\frac{Y}{4F}dw^{2}+\sum_{i=1}^{2}\left(w+q_{i}\right)\left(d\mu_{i}^{2}+\mu_{i}^{2}d\phi_{i}^{2}\right)+\frac{F-1}{Y}\left(\sum_{i=1}^{2}\mu_{i}^{2}d\phi_{i}\right)^{2}
\end{align}
with
\be
\sum_{i=1}^{2}\mu_{i}^{2}  =1,\qquad Y=\sum_{i=1}^{2}\frac{\mu_{i}^{2}}{w+q_{i}}
\ee
and $F$ an arbitrary function of $w$.
To show that the metric is K\"ahler we introduce the orthonormal frame
\begin{align}
e_{i}= & \frac{1}{2\sqrt{F}}\frac{\mu_{i}}{\sqrt{w+q_{i}}}dw+\sqrt{w+q_{i}}d\mu_{i}\nn
\bar{e}_{i}= & \frac{\sqrt{F}-1}{Y}\frac{\mu_{i}}{\sqrt{w+q_i}}\sum_{j=1}^{2}\mu_{j}^{2}d\phi_{j}+\sqrt{w+q_{i}}\mu_{i}\, d\phi_{i}
\end{align}
with
\be
ds_{4}^{2}=\sum_{i=1}^{2}\left(e_{i}\otimes e_{i}+\bar{e}_{i}\otimes\bar{e}_{i}\right).
\ee
The K\"ahler form can be written
\begin{align}
J= & \frac{i}{2}\sum_{i=1}^{2}\left(e_{i}-i\bar{e}_{i}\right)\wedge\left(e_{i}+i\bar{e}_{i}\right)=-\sum_{i=1}^{2}e_{i}\wedge\bar{e}_{i}\nn
= & -\frac{1}{2}dw\wedge\sum_{i=1}^{2}\mu_{i}^{2}d\phi_{i}-\sum_{i=1}^{2}\left(w+q_{i}\right)\mu_{i}d\mu_{i}\wedge d\phi_{i}\end{align}
which is clearly closed for any choice of $F$.

The holomorphic $(2,0)$-form $\Omega$ is given by
\begin{align}
\Omega= & \prod_{i=1}^{2}\left(e_{i}-i\bar{e}_{i}\right)\nn
= & \sqrt{w+q_{1}}\sqrt{w+q_{2}}\left[\frac{Y}{2\sqrt{F}}\, dw\wedge d\theta-\sqrt{F}\cos\theta\sin\theta\, d\phi_{1}\wedge d\phi_{2}\right]\nn
 & -i\sqrt{w+q_{1}}\sqrt{w+q_{2}}\frac{1}{2\sqrt{F}}\cos\theta\sin\theta\, dw\wedge\left(\frac{d\phi_{2}}{w+q_{1}}-\frac{d\phi_{1}}{w+q_{2}}\right)\nn
 & +i\sqrt{w+q_{1}}\sqrt{w+q_{2}}\sqrt{F}\, d\theta\wedge\left(\cos^{2}\theta\, d\phi_{1}+\sin^{2}\theta\, d\phi_{2}\right)\end{align}
where we have introduced $\mu_{1}=\cos\theta$, $\mu_2=\sin\theta$, $0<\theta<\frac{\pi}{2}$.
A calculation now shows that
\be
d\Omega=i P\wedge\Omega
\ee
with
\begin{align}
P= & \frac{2\sqrt{F}}{Y\sqrt{w+q_{1}}\sqrt{w+q_{2}}}\partial_{w}\left(\sqrt{F}\sqrt{w+q_{1}}\sqrt{w+q_{2}}\right)\,\left(\cos^{2}\theta\, d\phi_{1}+\sin^{2}\theta\, d\phi_{2}\right)\nn
 & +\frac{1}{Y}\cos2\theta\,\left(\frac{d\phi_{2}}{w+q_{1}}-\frac{d\phi_{1}}{w+q_{2}}\right).
\end{align}
From this we deduce that the complex structure is integrable, and thus we do indeed have a local K\"ahler metric with
Ricci form given by $dP$.
It is helpful to observe that we can also write
\begin{align}
\label{potk}
P
= & \partial_{w}\left[\left(F-1\right)\left(w+q_{1}\right)\left(w+q_{2}\right)\right]\frac{\sum_{i=1}^{2}\mu_{i}^{2}d\phi_{i}}{Y\left(w+q_{1}\right)\left(w+q_{2}\right)}
+d\phi_1+d\phi_2.
\end{align}

We now construct a closed two-form $W$ which satisfies
\be\label{oneone}
\Omega\wedge W=0,
\ee
which is the condition for it to be a $\left(1,1\right)$-form, and also
\be
J\wedge W=0,
\ee
which is the condition for it to be a primitive two-form. We make the ansatz
\be
W=d\left[f\left(w\right)\,\frac{\sum_{i=1}^{2}\mu_{i}^{2}d\phi_{i}}{Y\left(w+q_{1}\right)\left(w+q_{2}\right)}\right]
\ee
which satisfies the first equation. The second equation
reads
\be
J\wedge W=-\frac{\partial_{w}f}{Y\left(w+q_{1}\right)\left(w+q_{2}\right)}\, J\wedge J=0
\ee
and so we take
\be
W=Q\, d\left[\frac{\sum_{i=1}^{2}\mu_{i}^{2}d\phi_{i}}{Y\left(w+q_{1}\right)\left(w+q_{2}\right)}\right]
\ee
where $Q$ is a constant. The two-form $W$ is anti-self dual and we note that
\be
W^{ij}W_{ij}=\frac{16Q^{2}}{\left[Y\left(w+q_{1}\right)\left(w+q_{2}\right)\right]^{4}}.
\ee

Having fixed $W$, and hence the three-form flux $G$, we just need to fix the function $F$ to obtain the K\"ahler metric $ds^2_4$
by solving \reef{miib} which reads
\be\label{bigeq}
\Box R-\frac{1}{2}R^{2}+{R}^{ij}{R}_{ij}+4W^{ij}W_{ij}=0.
\ee
We consider the ansatz
\be
F=1+\lambda w^{2}\prod_{i=1}^{2}\frac{1}{w+q_{i}}+\Lambda\prod_{i=1}^{2}\frac{1}{w+q_{i}},
\ee
observing from \reef{potk} that the constant $\Lambda$ does not enter the
Ricci potential. A calculation shows that the Ricci scalar is given by
\be
R=-\frac{8\lambda}{Y\left(w+q_{1}\right)\left(w+q_{2}\right)}.
\ee
and that \reef{bigeq} boils down to solving
\begin{align}
\frac{\Lambda}{Y\left(w+q_{1}\right)\left(w+q_{2}\right)}\partial_{w}^{2}R+W^{ij}W_{ij} & =0
\end{align}
which implies that $\Lambda  =\frac{Q^{2}}{\lambda}$.

In summary, supersymmetric $AdS_3$ solutions of type IIB supergrvaity can be constructed from
the six-dimensional K\"ahler metric \reef{mbn}, with
the four-dimensional K\"ahler metric given by
\begin{align}
ds_{4}^{2} & =\frac{Y}{4F}dw^{2}+\sum_{i=1}^{2}\left(w+q_{i}\right)\left(d\mu_{i}^{2}+\mu_{i}^{2}d\phi_{i}^{2}\right)+\frac{F-1}{Y}\left(\sum_{i=1}^{2}\mu_{i}^{2}d\phi_{i}\right)^{2}
\end{align}
and
\be
F=1+\left(\lambda w^2+\frac{Q^{2}}{\lambda}\right)\frac{1}{\left(w+q_{1}\right)\left(w+q_{2}\right)}.
\ee
The three-form flux is given by \reef{gbn} with the closed, primitive and $\left(1,1\right)$-form $W$ given by
\be
W=Q\, d\left[\frac{\sum_{i=1}^{2}\mu_{i}^{2}d\phi_{i}}{Y\left(w+q_{1}\right)\left(w+q_{2}\right)}\right].
\ee

Observe that when $q_1=q_2\equiv q$, the metric is precisely of the form found in \cite{ajn} leading
to the $AdS_3$ solutions that we have analysed in detail in this paper.
To see this we let $w+q=1/x$ and we also introduce Euler angles via
\bea
\mu_1e^{i\phi_1}&=&\cos\frac{\theta}{2}e^{i\frac{\psi+\phi}{2}}\nn
\mu_2e^{i\phi_2}&=&\sin\frac{\theta}{2}e^{i\frac{\psi-\phi}{2}}.
\eea
We then find that
\begin{align}
ds^2_4=\frac{dx^2}{4x^3U}+\frac{1}{4x}(d\theta^2+\sin^2\theta d\phi^2)+\frac{U}{4x}(d\psi+\cos\theta d\phi)^2
\end{align}
with
\be
U=1+\lambda(1-qx)^2+\frac{Q^2}{\lambda}x^2
\ee
which should be compared with equations C.1 and C.7 of \cite{ajn}.
Furthermore,
\be
W=\frac{Q}{2}d[x(d\psi+\cos\theta d\phi)]
\ee
which should be compared with equation C.5 of \cite{ajn}.
When $q_1=q_2$, the metric $ds^2_4$ has local isometry group  $SU(2)\times U(1)$ and
the metric is cohomogeneity one. In the more general solutions with $q_1\ne q_2$ the
local isometry group is $U(1)\times U(1)$ and the metric is cohomogeneity two.

It will be interesting to analyse these more general $AdS_3$ solutions with $q_1\ne q_2$ in more detail.
When $Q=0$ the internal space will have topology $S^2\times S^3\times T^2$ and
when $Q\ne 0$ it will have topology $S^3\times S^3\times S^1$.
This can be shown using the techniques used in \cite{Cvetic:2005ft} and in this paper.
When $Q\ne 0$, one will also need to check the flux quantisation conditions and
this will require generalising the techniques that we have used in this paper. We leave this for the future.

\end{document}